\documentclass[twocolumn,notitlepage,footinbib,aps,prb,superscriptaddress]{revtex4-1}
\usepackage{graphicx,amsmath,bm,amssymb}
\begin{document}
\title{The phonon drag force acting on a mobile crystal defect:\\
full treatment of discreteness and non-linearity}
\author{T D Swinburne}
\email{thomas.swinburne@ccfe.ac.uk}
\affiliation{CCFE, Culham Science Centre, Abingdon, Oxon, OX14 3DB, UK}
\author{S L Dudarev}
\affiliation{CCFE, Culham Science Centre, Abingdon, Oxon, OX14 3DB, UK}
\date{\today}
\begin{abstract}
\noindent
Phonon scattering calculations predict the drag force acting on defects and dislocations rises linearly with temperature, in direct contradiction with molecular dynamics simulations that often finds the drag force to be independent of temperature. Using the Mori-Zwanzig projection technique, with no recourse to elasticity or scattering theories, we derive a general Langevin equation for a crystal defect, with full treatment of discreteness and non-linearity in the defect core. We obtain an analytical expression for the drag force that is evaluated in molecular statics and molecular dynamics, extracting the force on a defect directly from the inter-atomic forces. Our results show that a temperature independent drag force arises because vibrations in a discrete crystal are never independent of the defect motion, an implicit assumption in any phonon-based approach. This effect remains even when the Peierls barrier is effectively zero, invalidating qualitative explanations involving the radiation of phonons. We apply our methods to an interstitial defect in tungsten and solitons in the Frenkel-Kontorova model, finding very good agreement with trajectory-based estimations of the thermal drag force.
\end{abstract}
\maketitle
Crystalline materials are invariably host to a huge population of defects such as dislocation lines, dislocation loops, vacancies, impurities and self-interstitial atoms. As is well known, the defect dynamics are typically a non-inertial mixture of drift and diffusion due to significant interaction of defects with thermal vibrations\cite{kubin2013}. In this paper we describe in detail a recently reported method\cite{swinburne2014} for evaluating and understanding the interaction between thermal vibrations and crystal defects, resulting in a treatment of defect mobility that resolves an acute failing of phonon scattering theories.\\

The dynamical law used to describe defect motion typically balances deterministic forces from elastic interactions $\rm f$ against a viscous drag $-{\gamma}{\rm v}$, where ${\rm v}$ is the defect velocity (possibly that of a node on a dislocation line\cite{BulatovBook}), whilst $\gamma$ is the drag coefficient, giving ${\rm v}={\rm f}/\gamma$. In dislocation dynamics literature\cite{BulatovBook} the drag coefficient $\gamma$, sometimes labelled by $B$, is related to the mobility $M=1/\gamma$, so that $\dot{\rm x}=M{\rm f}$; in this paper we will use $\gamma$ throughout. A common issue with this purely viscous dynamical law is the absence of any temperature and thus thermal fluctuations. It has recently been shown\cite{Dudarev2011,swinburne2013} that to correctly capture the highly stochastic trajectories of nano-scale defects and kink-bearing dislocation lines seen in experiment\cite{Arakawa2007,yi2013} one must also add a stochastic thermal force ${\eta}(t)$ to the defect equation of motion, giving the Langevin equation
\begin{equation}
{\rm v}={\rm f}/\gamma+{\eta}(t)/\gamma,\label{INTRO_LOD}
\end{equation}
where $\eta(t)$ is a white noise defined by the ensemble averages\cite{coffey2004}
\begin{equation}
\langle{\eta}(t)\rangle={0},\quad\langle{\eta}(t){\eta}(t')\rangle=2{\rm k_BT}{\gamma}\delta(t-t'),\label{INTRO_STOCH}
\end{equation}
derived from the fluctuation-dissipation theorem\cite{reif}. The stochastic thermal force ${\eta}$ is particularly relevant for nanoscale defects and small dislocation loops as they only respond to stress gradients\cite{Hirth}, meaning that the elastic force $\rm f$ is often negligible and the stochastic force dominates, giving the purely diffusive equation of motion $\dot{\rm x}={\eta}(t)/\gamma$ with a diffusion constant $D=\lim_{t\to\infty}\langle x^2(t)\rangle/2t={\rm k_BT}/\gamma$.  For long dislocation lines the elastic force $f$ typically dominates, giving an expected velocity of $\langle\dot{x}\rangle=\lim_{t\to\infty}\langle x(t)\rangle/t={\rm f}/\gamma$, as the expected stochastic force $\langle\eta\rangle=0$ vanishes. In either case, the value of $\gamma$ sets the time scale of defect motion and defines the dynamics of ensembles of interacting defects, and thus is a critical parameter in any dislocation mediated process such as post-irradiation annealing of defects, plastic deformation or the ductile-brittle transition.\\

Whilst significant effort has gone into evaluating the deterministic elastic forces acting on crystal defects, there has been relatively little theoretical effort focused on the evaluation of the drag parameters $\gamma$ that play such a crucial role in any simulation of defect dynamics. All theoretical estimates use the results of Nabarro\cite{Nabarro}, Eshelby\cite{eshelby1962} and others\cite{brailsford1970,alshits1975}, who calculated dissipation rates due to scattering by thermal phonons. In such treatments the drag force $-\gamma{\rm v}$ is calculated to be proportional to the phonon density, meaning the drag coefficient is predicted to rise at least linearly with temperature in the classical limit, i.e. $\gamma\simeq\gamma_{\rm w}{\rm k_BT}$. This so-called `phonon wind' relationship is universally invoked, but has at best partial, qualitative, agreement with molecular dynamics (MD) simulation, agreeing with some simulations of dislocation lines in FCC metals\cite{bitzek2004,bitzek2005} but not identical simulations in BCC metals\cite{Chang,Gilbert2011,Gilbert,kuksin2013,dislolennote}, where the analysis of the average velocity under stress reveals $\gamma=f/\langle\dot{x}\rangle=\gamma_0+\gamma_{\rm w}{\rm k_BT}$, where $\gamma_0$ is independent of temperature. Most dramatically, simulations of nanoscale dislocation loops\cite{Dudarev2011,Dudarev2008b}, screw dislocation kinks\cite{Gilbert,swinburne2013} and highly mobile self-interstitial crowdion defects\cite{zepeda2005} in BCC metals, reveal a diffusivity that rises linearly with temperature, meaning $\gamma={\rm k_BT}/D=\gamma_0$ is independent of temperature. This surprising result is in complete disagreement with the phonon wind theory, which as we show below explicitly forbids the existence of a temperature independent component $\gamma_0$. Qualitative explanations for $\gamma_0$ claim dissipation arises due to phonon radiation as the defect passes over the Peierls migration barrier\cite{alshits1975,kuksin2013} lack any quantitative foundation and can clearly not account for the presence of $\gamma_0$ when the Peierls barrier is extremely small, as is the case for e.g. kinks on screw dislocations\cite{Ventelon2009} and crowdions\cite{Fitzgerald2008} in BCC metals.\\

The lack of any quantitative theory for the presence of a temperature independent drag force highlights a need for a fundamental, quantitative understanding of the interaction of crystal defects with thermal lattice vibrations. In this paper we detail a recently reported\cite{swinburne2014} treatment of the defect drag force that fully accounts for the non-linear, discrete character of a real crystal and does not rely on elasticity and phonon scattering theories. We introduce defects as general localized deformations in the atomic configuration of a simulated crystal (section \ref{sec:PO}), deriving a stochastic equation of defect motion using the established Mori-Zwanzig technique\cite{Zwanzig} (\ref{sec:ZW}). It is shown that in appropriate limits the drag coefficient $\gamma$ is equal to the integrated time autocorrelation of the defect force, divided by temperature, a Green-Kubo relation for the defect force. By taking averages over a quadratic Gibbs distribution of appropriately defined fluctuations defined in the subspace orthogonal to coordinates of the moving defect itself, we derive an analytic expression for the autocorrelation of the defect force (\ref{sec:AG}) and find that in general
\begin{equation}
\gamma=\gamma_0+{\rm k_BT}\gamma_{\rm w},
\end{equation}
where $\gamma_0$ and $\gamma_{\rm w}$ are constants. This central result provides the a quantitative, theoretical demonstration of a temperature independent drag force for a general crystal defect with full treatment of non-linearity and discreteness, explaining the findings of many independent simulation studies\cite{swinburne2013,Gilbert,Dudarev2008b,zepeda2005,kuksin2013}.

We analyse the form of our explicit analytical expression for $\gamma_0$ (section \ref{sec:AG}), showing that the vibrations orthogonal to the defect motion are in general not vibrational modes of the crystal, by which we mean that vibrational displacements orthogonal to the defect motion produce a force that has a component  parallel to the defect motion, even to linear order in the vibrational displacements. This means that phonons are not perfect oscillators perturbed by anharmonic couplings and consequently phonon momentum cannot be conserved in a defective crystal, invalidating the results of any phonon-defect scattering theory. These conclusions are explicitly tested on the discrete Frenkel-Kontorova model\cite{swinburne2013b} (\ref{sec:FK_I}), a very special case as the continuum limit is an integrable system, with the `defect' becoming a soliton\cite{March1987}. In this special case $\gamma_0$ is predicted to vanish in the continuum limit, which we test in numerical simulations. Although the Peierls barrier is observed to vanish as we approach the continuum limit, we find that $\gamma_0$ remains, as fluctuations in a discrete lattice are always distinct from those in the continuum. This result shows that the presence of $\gamma_0$ is not related to any migration barrier effects, further invalidating a qualitative explanation of $\gamma_0$ based on phonon radiation\cite{alshits1975,kuksin2013}. In addition, the fact that $\gamma_0$ remains even in such an idealised system is a testament to the generality of our results.\\

In the final section (\ref{sec:NI}), we apply our approach to a realistic crystal defect, the $1/2\langle111\rangle$ interstitial crowdion in Tungsten. Our approach offers a clear method to calculate the force acting on a defect directly from the inter-atomic forces, allowing a force autocorrelation and thus $\gamma$ to be calculated in molecular dynamics simulation runs and also by evaluating our analytical expression in molecular statics, finding very good agreement with typical measurements of $\gamma$ from diffusive simulation trajectories. It is hoped the present work will allow a better understanding of dissipation in crystalline systems and highlights the extent to which real crystal defects are fundamentally discrete and anharmonic objects that can never be assigned a conserved energy and momentum in a bath of canonical phonons.\\

\section{A qualitative summary of classical phonon scattering theory}\label{sec:PS_INTRO}
Previous attempts to calculate an effective drag force on crystal defects all aim to calculate the scattering cross-section that defects offer to thermal phonons. Whilst we will not go into a detailed derivation of these calculations\cite{Nabarro,brailsford1970,alshits1975,Argon} we provide a qualitative summary of how phonon scattering theory forbids a temperature independent drag parameter, i.e. $\gamma_0=0$, in conflict with extensive MD simulation results. \\
\begin{figure}
\centering\includegraphics[width=0.45\textwidth]{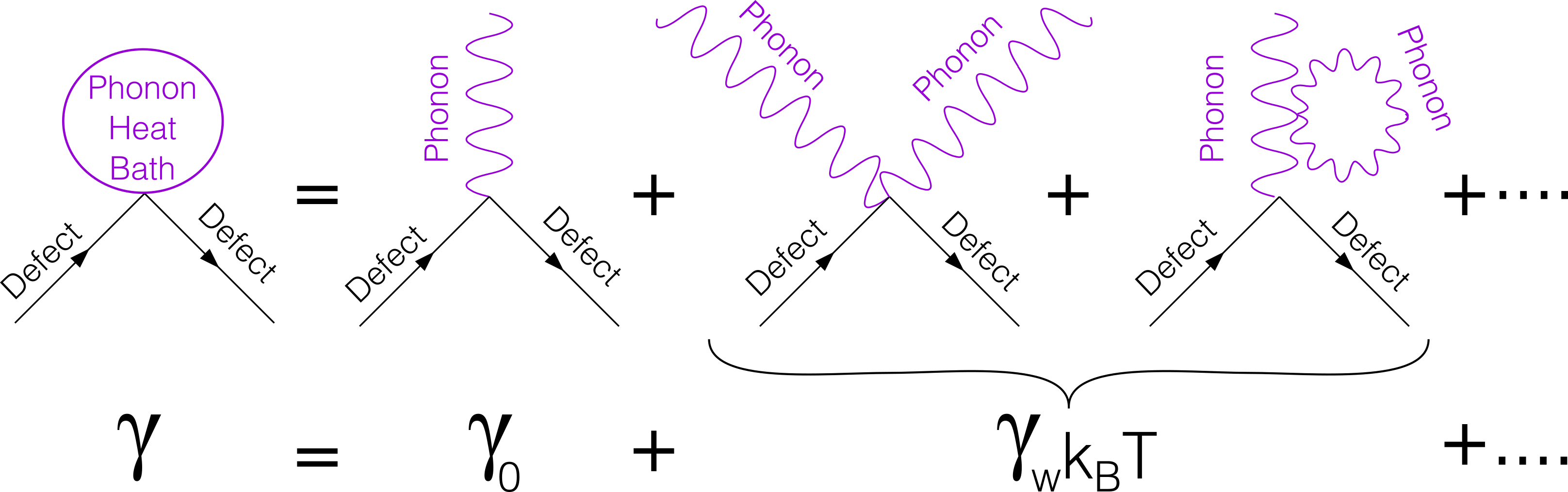}
\caption{A cartoon of phonon scattering by a dislocation. Higher order scattering processes contribute terms of higher order in temperature to the effective drag parameter $\gamma$. As demonstrated in the text, phase space limitations result in the first order term $\gamma_0=0$ vanishing for subsonic defect speeds. The second diagram represents a typical scattering process which leads to the well-known\cite{Nabarro,eshelby1962,alshits1975} `Phonon Wind' relationship $\gamma=\gamma_{\rm w}{\rm k_BT}$.\label{ps_diag}}
\end{figure}
A founding assumption of phonon scattering theory is that defects and phonons can each be considered initially free particles, with quadratic and linear dispersion relations respectively. Drag occurs due to interactions through well defined scattering processes. In such a picture, the drag parameter $\gamma=\gamma_0+\gamma_{\rm w}{\rm k_BT}+\gamma_2({\rm k_BT})^2+...$ represents a thermally averaged cross section to all scattering processes. As the phonon density is proportional to ${\rm k_BT}$ in the classical limit, higher order terms in $\gamma$ will be the result of scattering processes involving a greater number of phonons. It turns out\cite{Nabarro,alshits1975} that the terms of order $({\rm k_BT})^n$ involve $n+1$ phonons, as illustrated in Figure \ref{ps_diag}. For $\gamma_0$ the scattering process is therefore particularly simple- the absorption or emission of a single phonon.\\

To see why a phonon scattering approach invariably predicts that $\gamma_0=0$, consider a crystal with acoustic wave speed $c$, lattice constant $a$ and atomic mass $m$. Phonons have an energy $E=|p|c$ and momentum $|p|\leq h/a\ll mc$, whilst dislocations have an energy $E=P^2/2{\tilde m}$ and momentum $P$, where $\tilde{m}\gtrsim m/5$ is the dislocations' effective unit core cell mass. As we are far from any shock front and under typical mechanical stresses the dislocation speed is subsonic (far below the wave speed) i.e. $P\ll {\tilde m}c$. With a final dislocation momentum of $P'=P+p$ the energy balance thus reads
\begin{equation}
\frac{P^2}{2{\tilde m}}+|p|c=\frac{(P+p)^2}{2{\tilde m}},\Rightarrow\,P={\tilde m}c+p/2\simeq{\tilde m}c,
\end{equation}
which clearly violates the requirement that $P\ll {\tilde m}c$. As a result, the phase space available for such a one phonon scattering process vanishes and we are forced to conclude that $\gamma_0=0$, meaning $\gamma=\gamma_1{\rm k_BT}+\gamma_2{\rm k_BT}+...$ must rise at least linearly with temperature. This very reasonable argument has been invoked by all authors investigating the scattering of defects by thermal phonons\cite{eshelby1962,alshits1975,Dudarev2002}, but unfortunately is in conflict with many MD simulations of defect motion, as mentioned above.\\

We have seen that the conclusion $\gamma_0=0$ is a direct consequence of assuming defects are canonical objects- initially freely moving particles that interact via scattering processes, with a well defined and conserved energy and momentum. In the following sections we show that this founding assumption is always false as defect motion is fundamentally discrete and anharmonic. Also, the vibrational modes of the crystal always change with defect position, meaning phonons cannot be defined as canonical objects with conserved energy and momentum, invalidating all the calculations of phonon scattering theory. Instead, to obtain a correct expression for the defect drag force we must treat the defect position and velocity simply as functions of the full crystal configuration, deriving a general equation of motion using the established coarse graining techniques of the Mori-Zwanzig formalism\cite{Zwanzig}.
\section{Deriving an equation of defect motion}\label{sec:PO}
\subsection{Defects in a discrete crystal}
The state of a classical crystal of $N$ atoms is entirely defined by the $N$ atomic positions ${\bf x}_i\in\mathbb{R}^3$ and velocities $\dot{\bf x}_i\in\mathbb{R}^3$, where $i\in[1,N]$. It will be convenient to represent these positions and velocities by two 3N-dimensional vectors ${\bf X}, \dot{\bf X} \in \mathbb{R}^{\rm 3N}$, constructed from the atomic positions and velocities with the tensor sum
\begin{equation}
{\bf X}\equiv\left({\bf x}_1,{\bf x}_2,...,{\bf x}_{\rm N}\right),
\quad\dot{\bf X}\equiv\left(\dot{\bf x}_1,\dot{\bf x}_2,...,\dot{\bf x}_{\rm N}\right).
\end{equation}
Molecular dynamics algorithms assign a potential energy $V({\bf X})$ for the system and then integrate Newton's equation
${m}\ddot{\bf X}=-{\bm\nabla}V({\bf X})$ under thermodynamic constraints to generate classical dynamics. It is clear that this inherently discrete system cannot exhibit the elastic singularities that form defects in a continuum system. Instead, defects are localized deformations which can be described by assigning a set of M$\;\ll\;$N nodes, with positions ${\bf r} \in \mathbb{R}^{\rm M}$ and velocities $\dot{\bf r}={\bf v} \in \mathbb{R}^{\rm M}$. We emphasize that ${\bf r},{\bf v}$ have no associated equation of motion, unlike the position and velocity of a particular atom. The dynamics of these nodes will be a projection of the Newtonian dynamics of the whole crystal. Many methods exist for determining the defect position and velocity, including analysis of the centrosymmetry parameter\cite{BulatovBook}, Voronoi analysis\cite{stukowski2010} and finding peaks in the time averaged potential energy\cite{swinburne2013}. In the following we do not constrain ourselves by the choice of a particular method for the identification of the defect and only require that any method used is consistent and repeatable.\\

The number of nodes one should assign to a crystal defect will obviously increase with the defect size. For example, to fully capture the configurational complexity of an extended dislocation line it is necessary to assign a node to each atomic plane normal to the dislocation line direction\cite{swinburne2013}. However, in the present work we focus, in the interest of clarity, on very simple defects which can be represented by a single node that only moves along a single direction, allowing us to set $\rm M=1$, such that ${\bf r}={\rm r}\in\mathbb{R}$ and ${\bf v}={\rm v}\in\mathbb{R}$.\\
\begin{figure}
\centering\includegraphics[width=0.45\textwidth]{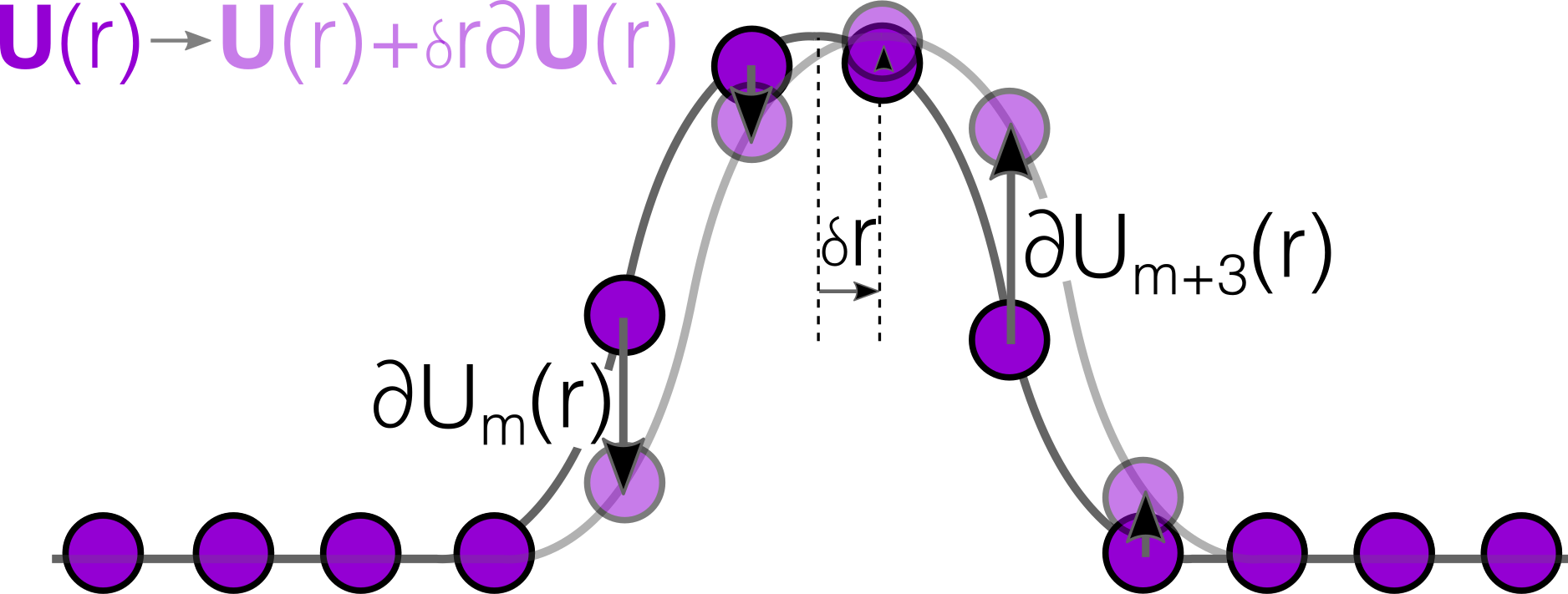}
\caption{An illustration of the defect translation vector $\partial_{\rm r}{\bf U}$ for a localised `hump' in a chain of `atoms'. The vector $\partial_{\rm r}{\bf U}$ describes the individual atomic displacements that correspond to an infinitessimal defect migration at zero temperature.\label{hump}}
\end{figure}
At zero temperature, there exists a well defined atomic configuration ${\bf X}$=${\bf U}({\rm r})$ that minimizes the potential energy of the crystal $V({\bf X})$ for each value of defect position ${\rm r}$. These configurations are what is found in, for example, nudged elastic band calculations\cite{neb}. At a finite temperature, these minimum energy configurations will be augmented by the displacements due to thermal vibrations $\bm\Phi \in \mathbb{R}^{\rm 3N}$, meaning the crystal configuration $\bf X$ at any given instant can be expressed as
\begin{equation}
{\bf X} = {\bm\Phi} + {\bf U}({\rm r})
\,,\,
\dot{\bf X} = \dot{\bm\Phi}+{{\rm v}}\partial_{\rm r}{\bf U},
\label{first}
\end{equation}
where $\partial_{\rm r}$ is the partial derivative with respect to ${\rm r}$ whilst keeping $\bm\Phi$ constant.
For the $3N$-dimensional vector ${\bf U}({\rm r})$, which depends only on a single variable $\rm r$, $\partial_{\rm r}$ is equivalent to simple ordinary differentiation and can be readily evaluated. In the general case, we use the differential operator $\bm\nabla={\partial}/{\partial\bf X}$, to define $\partial_{\rm r}$ through its action on a (possibly vectorial) function ${\bf A}({\bf X})$ as
\begin{align}
\partial_{\rm r}{\bf A}({\bf X})&\equiv\lim_{\delta\to0}\frac{{\bf A}({\bf U}({\rm r+\delta})+{\bm\Phi})-{\bf A}({\bf U}({\rm r})+{\bm\Phi})}{\delta}\nonumber\\
&\equiv\left(\frac{\partial {\bf A}}{\partial{\rm r}}\right)_{\bm\Phi}\equiv\left({\partial_{\rm r}}{\bf U}\cdot{\bm\nabla}\right){\bf A},\label{P_dx}
\end{align}
which is clearly the infinitesimal change in $\bf A$ due to motion along $\partial_{\rm r}{\bf U}$. As illustrated in figure \ref{hump}, the vector $\partial_{\rm r}{\bf U}$ is very important as it defines the directions of $3N$ atomic displacements necessary for defect migration and thus singles out the defect dynamics from the dynamics of the crystal as a whole.
The mathematical treatment of $\partial_{\rm r}{\bf U}$ can be introduced using the trivial case of the defect being a physical atom. Assigning an index $j$ to the atom, the 3N-dimensional vector $\partial_{\rm r}{\bf U}$ consistent with the variation of, say, the $y$-coordinate $y_j$ of atom $j$, is simply $\partial_{y_j}{\bf U}=(0,0...1,0...0)\in\mathbb{R}^{3\rm N}$. This vector picks out the relevant atomic coordinate from a full crystal configuration. All the components of vector $(0,0...1,0...0)$ are zeros, apart from the one component in $3j-1$ position, which is equal to 1. The set of the $3N-1$ other directions orthogonal to $(0,0...1,0...0)$ in this case are defined by the $3N \times 3N$ matrix ${\rm Diag}(1,1...0,1...1)$, where zero is in $3j-1$ position. The action of this matrix on an arbitrary $3N$-dimensional vector generates a vector that is orthogonal to $(0,0...1,0...0)$.

For a localised deformation such as a crystal defect, the vector ${\rm\partial}_{\rm r}{\bf U}$ now defines a direction in the $3N$-dimensional space of atomic coordinates associated with the motion of a defect from $r$ to $r+\delta r$, as illustrated in figure \ref{hump}. The matrix defining the space of all directions orthogonal to ${\rm\partial}_{\rm r}{\bf U}$ is now given by
\begin{equation}
{\mathbb{I}}-\frac{\partial_{\rm r}{\bf U}\otimes\partial_{\rm r}{\bf U}}{\partial_{\rm r}{\bf U}\cdot\partial_{\rm r}{\bf U}}.\label{Q_M}
\end{equation}
To illustrate the point, consider the action of this matrix on an arbitrarily chosen vector ${\bf Y}$,
\begin{equation}
{\bf Z}=\left[\mathbb{I}-\frac{\partial_{\rm r}{\bf U}\otimes\partial_{\rm r}{\bf U}}{\partial_{\rm r}{\bf U}\cdot\partial_{\rm r}{\bf U}}\right]\cdot{\bf Y}={\bf Y}-\frac{\partial_{\rm r}{\bf U}\cdot {\bf Y}}{\partial_{\rm r}{\bf U}\cdot \partial_{\rm r}{\bf U}}\partial_{\rm r}{\bf U}.
\end{equation}
Clearly, ${\bf Z}$ is orthogonal to $\partial_{\rm r}{\bf U}$ since
\begin{equation}
\partial_{\rm r}{\bf U}\cdot {\bf Z}=\partial_{\rm r}{\bf U}\cdot {\bf Y}-\frac{\partial_{\rm r}{\bf U}\cdot \partial_{\rm r}{\bf U}}{\partial_{\rm r}{\bf U}\cdot \partial_{\rm r}{\bf U}}\left( \partial_{\rm r}{\bf U}\cdot {\bf Y}\right)=0.
\end{equation}

To see how this relates to the description of a crystal at finite temperature, we first note that the space of variables including phonon displacements, velocities, and a single defect position and velocity ${\bm\Phi}\oplus\dot{\bm\Phi}\oplus{\rm r}\oplus{\rm v}$ has two more dimensions, i.e. $6N+2$, than the $6N$-dimensional space of atomic coordinates and velocities ${\bf X}\oplus\dot{\bf X}$. In order to ensure the same number of degrees of freedom in both coordinate sets we require the vibrational displacements $\bm\Phi$ to be orthogonal to the direction of displacements $\partial_{\rm r}{\bf U}$ caused by defect motion. This defines a 3N-1 dimensional hypersurface of thermal displacements\cite{March,Boesch1988,yip1984}
\begin{equation}
\partial_{\rm r}{\bf U}\cdot{\bm\Phi}=0.\label{cond_x}
\end{equation}
The constraint (\ref{cond_x}) on the thermal vibrations $\bm\Phi$ can be derived by varying defect position ${\rm r}$ to minimize the quadratic deviation $|{\bf X}-{\bf U}({\rm r})|^2$, giving the minimum condition $\partial_{\rm r}{\bf U}\cdot({\bf X}-{\bf U})=\partial_{\rm r}{\bf U}\cdot{\bm\Phi}=0$. For a given value of defect position $\rm r$, we can now vary ${\rm v}=\dot {\rm r}$ to minimize the quadratic deviation $|\dot{\bf X}-{\rm v}\partial_{\rm r}{\bf U}|^2$, resulting in $\delta |\dot{\bf X}-{\rm v}\partial_{\rm r}{\bf U}|^2=\delta {\rm v}[\partial_{\rm r}{\bf U}\cdot ( \dot{\bf X}-{\rm v}\partial_{\rm r}{\bf U})]=0$, which in combination with the second of equations (\ref{first}) gives
\begin{equation}
\partial_{\rm r}{\bf U}\cdot\dot{\bm\Phi}=0.\label{cond_v}
\end{equation}
The constraints (\ref{cond_x}) and (\ref{cond_v}) emphasize that by assigning a defect position and velocity to the crystal configuration, the 3N-dimensional vectors ${\bm\Phi},\dot{\bm\Phi}$ describing the vibrational displacements of each atom in the system are restricted to lie on the 3N-1 dimensional hypersurfaces defined by (\ref{cond_x}) and (\ref{cond_v})\footnote{We note that for the case of M defect co-ordinates we will have 2M constraints corresponding to each of the M partial derivatives of ${\bf U}$, meaning $\bm\Phi$ will account for $3N-M$ vibrational degrees of freedom\cite{swinburne2014}}. As each hypersurface will change as the defect moves (as $\partial_{\rm r}{\bf U}$ changes) the vibrational co-ordinates retain a dependence on ${\rm r}$, although this dependence on $\rm r$ does not feature in the differential operator $\partial_{\rm r}$ defined in (\ref{P_dx}) as $\bm\Phi$ is held constant. The representation of ${\bm\Phi}$ and $\dot{\bm\Phi}$ will be analysed in more detail when we calculate vibrational expectation values.\\

To obtain a defect dynamical equation, we project the conventional atomic equations of motion $m\ddot{\bf X} = -{\bm\nabla}V(\bf X)$ onto the vector $\partial_{\rm r}{\bf U}$ defining the direction of defect motion. Forming a scalar product of the equations of motion with vector $\partial_{\rm r}{\bf U}$, we obtain $m\partial_{\rm r}{\bf U}\cdot\ddot{\bf X} = -\partial_{\rm r}{\bf U}\cdot{\bm\nabla}V(\bf X)$. We emphasize that as $\rm r$ is not a canonical co-ordinate we do not expect that the equation of motion for the defect to resemble a Hamiltonian equation of motion; we aim to derive from the true equations of motion $m\ddot{\bf X} = -{\bm\nabla}V(\bf X)$ for the atoms an expression that defines the dynamical law that defect position $\rm r$ obeys. Differentiating the second of equations (\ref{first}) with respect to time, and noting that $\dot {\rm r}={\rm v}$, we find that $\ddot{\bf X}=\ddot{\bm\Phi}+{{\rm v}^2}{\partial}^2_{\rm r}{\bf U}+\dot{\rm v}\partial_{\rm r}{\bf U}$. Exploiting the constraints (\ref{cond_x}),(\ref{cond_v}) we arrive at
\begin{equation}
\tilde{m}\dot{\rm v} =
-\partial_{\rm r}{\bf U}\cdot{\bm\nabla}V({\bf X}) - (\partial_{\rm r}\tilde{m}){\rm v}^2/2
+{\rm v}\cdot\partial^2_\lambda{\bf U}\cdot\dot{\bm\Phi},
\end{equation}
where $\tilde{m}= m\partial_{\rm r}{\bf U}\cdot\partial_{\rm r}{\bf U}$ is the effective mass of the defect and $\partial_{\rm r}\tilde{m}=2m\partial^2_{\rm r}{\bf U}\cdot\partial_{\rm r}{\bf U}$. Similar equations are known in other dynamical quasiparticle theories\cite{March1987,Boesch1988}. Following the established approximations, we neglect the `hydrodynamic' term $-{\rm v}\partial^2_\lambda{\bf U}\cdot\dot{\bm\Phi}$ and the effective kinetic energy gradient $-\partial_{\rm r}\tilde{m}{\rm v}^2/2$. This is justified as we are only considering defects with subsonic speeds $|{\rm v}|\ll c$ and small migration barriers.  For the effective kinetic energy gradient, as the defect core is wide when the migration barrier is small (see below), the effective mass is of order $1/c$\cite{Braun1998} and varies very little with position, meaning $-\partial_{\rm r}\tilde{m}{\rm v}^2/2$ is (at most) of order $({\rm v}/c)^2\ll1$. The `hydrodynamic' term $-{\rm v}\partial^2_\lambda{\bf U}\cdot\dot{\bm\Phi}$ will have a vanishing expectation value as $\langle\dot{\bm\Phi}\rangle=0$ (we will define this expectation value precisely below), which in turn means that this term will only appear as second order or higher, which for a wide defect core will also give a contribution of order $({\rm v}/c)^2\ll1$. Neglecting these terms gives a final dynamical equation for the defect coordinates of
\begin{equation}
\tilde{m}\dot{\rm v} =-{\partial_{\rm r}}V \equiv -\partial_{\rm r}{\bf U}\cdot{\bm\nabla}V({\bf X}).\label{DEF_EOM}
\end{equation}
In appendix \ref{app:esh} we show that in the absence of the thermal vibrations and under a weak applied stress, the force on a crystal defect given by (\ref{DEF_EOM}) is identical that famously derived by Eshelby\cite{eshelby1951}. However, the main results of this paper concern the interaction of crystal defects with thermal vibrations; in order to perform the analytical manipulation in section \ref{sec:AG} we also require dynamical equations for the vibrational coordinates. In the same spirit as above, we analyse the full equation of motion in the subspace of all directions orthogonal to $\partial_{\rm r}{\bf U}$, defined through the matrix (\ref{Q_M}), arriving at
\begin{equation}
{m}\ddot{\bm\Phi}=-\left[
\mathbb{I}-
\frac{\partial_{\rm r}{\bf U}\otimes\partial_{\rm r}{\bf U}}{\left(\partial_{\rm r}{\bf U}\cdot\partial_{\rm r}{\bf U}\right)}
\right]
\cdot{\bm\nabla}V\equiv-{\bm\nabla}_{\bm\Phi}V.\label{VIB_EOM}
\end{equation}
where we have introduced the differential operator in the subspace of vibrations ${\bm\nabla}_{\bm\Phi}$. As for $\partial_{\rm r}$ in (\ref{P_dx}), ${\bm\nabla}_{\bm\Phi}$ can be defined through its action on a function ${\bf A}$ as
\begin{equation}
{\bm\nabla}_{\bm\Phi}{\bf A}\equiv\left(\frac{\partial{\bf A}}{\partial{\bm\Phi}}\right)_{\rm r}\equiv
\left[
\mathbb{I}-
\frac{\partial_{\rm r}{\bf U}\otimes\partial_{\rm r}{\bf U}}{\left(\partial_{\rm r}{\bf U}\cdot\partial_{\rm r}{\bf U}\right)}
\right]
\cdot{\bm\nabla}{\bf A}.\label{VIB_D}
\end{equation}
The two dynamical equations (\ref{DEF_EOM}) and (\ref{VIB_EOM}) clearly show a close parallel with the standard classical equations of motion $m\ddot{\bf X}=-{\bm\nabla}V(\bf{X})$. However, the differential operators (\ref{P_dx}) and (\ref{VIB_D}) are defined through their relation to the direction of defect motion $\partial_{\rm r}{\bf U}$ in the 3N dimensional space of crystal configurations rather than the differentiation of a function with respect to a coordinate.  Whilst we have now derived a general dynamical equation for the defect co-ordinates (\ref{DEF_EOM}), they still retain an explicit dependence on the vibrational degrees of freedom. The next section uses the Mori-Zwanzig projection technique to replace the vibrational coordinates by a statistical distribution, producing a closed but stochastic equation of motion for the defect.
\subsection{Removing the vibrational coordinates by the Mori-Zwanzig method}\label{sec:ZW}
From the form of the equation of motion for the defect (\ref{DEF_EOM}) it is clear that the potential energy $V({\bf U}({\rm r})+{\bm\Phi})$ couples the evolution of the coordinate of the defect and vibrational coordinates.  This is what is required for the frictional force to exist. The general dynamic relationship between these coordinates is therefore highly complex, but in the present work we only consider defects with low migration barriers moving at subsonic speeds. This allows us to assume that the defect coordinates may be considered as slowly varying compared to the vibrational coordinates, an approximation the validity of which we will explicitly prove later when we calculate the defect force autocorrelation in MD simulations.\\

It is well known that defect migration barriers are directly related to the width of the defect core\cite{PeierlsOld}. The wider is the defect core the lower is the migration barrier. This is because if a defect possesses a wide core, defect motion induces only small individual atomic displacements; in Peierls' seminal paper\cite{PeierlsOld} and many subsequent treatments\cite{Joos1997,swinburne2013} it has been shown that defect migration barriers decay exponentially fast with the defect core size. As here we consider highly mobile subsonic defects with very small migration barriers, we are therefore only concerned with wide defect cores, exhibiting broad, in comparison with the lattice parameter, maxima in $\left(\partial_{\rm r}{\bf U}\right)^2$. We now use this fact to provide a heuristic argument justifying the assumed timescale separation. At a finite temperature, over a Debye period $\tau_{\rm D}$$\;\sim\;$$a/c$$\;\sim\;$$\;0.1$ps, where $a$ is the lattice parameter, the displacements of any atom due to thermal vibrations have the oscillation amplitude of $\sim\tau_{\rm D}\sqrt{{\rm k_BT}/{m}}$. Since the defect speed is approximately ${\rm v}\sim\sqrt{{\rm k_BT}/\tilde{m}}\ll c$, the displacement of any one atom {\it due to defect motion} over a time interval $\tau_{\rm D}$ will be at most $\tau_{\rm D}\|\partial_{\rm r}{\mathbf U}\|_\infty\sqrt{{\rm k_BT}/\tilde{m}}$, where $\|\partial_{\rm r}{\mathbf U}\|_\infty$ is the component of the greatest magnitude in $\partial_{\rm r}{\mathbf U}$. These calculations imply that if $\|\partial_{\rm r}{\mathbf U}\|_\infty\ll|\partial_{\rm r}{\mathbf U}|$, then the displacement due to defect motion will be much less than the magnitude of displacements due to thermal vibrations of atoms, which implies that the $\bm\Phi$ are effectively ergodic over a time-scale $\sim$ $\tau_{\rm D}$ where the defects essentially remains stationary. We note that the condition $\|\partial_{\rm r}{\mathbf U}\|_\infty\ll|\partial_{\rm r}{\mathbf U}|$ amounts to a requirement that the deformation associated with the defect is spread over many atomic sites, which is always satisfied by highly mobile defects with a wide core. We therefore assume that vibrational displacements average to zero over periods of $\sim$0.1ps whilst the displacements due to the defect structure are effectively static. Again, we will test this conclusion in MD simulation when calculating the defect force autocorrelation.\\

This separation of time-scales can be exploited to integrate out thermal vibrations from the defect equation of motion using the Mori-Zwanzig projection technique\cite{Zwanzig,chorin2000}. The idea is to derive a formal solution for the `fast' co-ordinates $\bm\Phi,\dot{\bm\Phi}$ which may be substituted into the equation for the `slow' defect co-ordinates ${\rm r},{\rm v}$. Crucially, by considering a distribution of initial conditions for the fast variables, we go from the micro canonical to canonical ensemble, introducing heat and stochastic fluctuations by only retaining statistical knowledge of the system. To actually do this we use a projection operator. The projection of some function $A({\rm r}(t),{\rm v}(t),{\bm\Phi}(t),\dot{\bm\Phi}(t))$ is a conditional average over the fast variables weighted by some probability distribution function $\rho$; for systems with a well defined temperature (meaning the vibrations are to leading order harmonic\cite{reif}) $\rho$ is simply the conditional Gibbs distribution, given by
\begin{equation}
\rho({\rm r},{\bm\Phi},\dot{\bm\Phi}) =  \exp\left({-\beta\left[V\left({\rm r},{\bm\Phi}\right)+ \frac{m}{2}\dot{\bm\Phi}\cdot\dot{\bm\Phi}\right]}\right)/Z({\rm r}),\label{FE}
\end{equation}
with the partial partition function $Z({\rm r})$ providing normalization. As defect velocity $\rm v$ only appears in the kinetic energy $\tilde{m}{\rm v}^2/2 +m\dot{\bm\Phi}\cdot\dot{\bm\Phi}/2$, it does not couple directly to the vibrational coordinates and so does not appear in (\ref{FE}). As discussed above, 
although the vibrational displacements ${\bm\Phi}$ and $\dot{\bm\Phi}$ appear as 3N-dimensional vectors, the constraints (\ref{cond_x}) and (\ref{cond_v}) restrict their allowed values to a 3N-1 dimensional hypersurface orthogonal to $\partial_{\rm r}{\bf U}$. Any conditional average over ${\bm\Phi}$, $\dot{\bm\Phi}$ must therefore be taken on this hypersurface, resulting in a projection operator 
\begin{align}
\hat{P}A(t)\equiv\int &A ({\rm r}(t),{\rm v}(t),{\bm\Phi},\dot{\bm\Phi})\rho({\rm r}(t),{\bm\Phi},\dot{\bm\Phi})\nonumber\\
&\times\delta(\partial_{\rm r}{\bf U}\cdot{\bm\Phi})\delta(\partial_{\rm r}{\bf U}\cdot\dot{\bm\Phi}){\rm d}{\bm\Phi}{\rm d}\dot{\bm\Phi},\label{proj_int}
\end{align}
where the delta functions signify that we integrate over the hyper-surfaces defined by the constraints (\ref{cond_x}) and (\ref{cond_v}). The normalization condition on $\rho$ is
\begin{align}
Z({\rm r})=\int
&e^{-\beta\left[V\left({\rm r},{\bm\Phi}\right)+m\dot{\bm\Phi}\cdot\dot{\bm\Phi}/2\right]}\nonumber\\
&\times\delta(\partial_{\rm r}{\bf U}\cdot{\bm\Phi})\delta(\partial_{\rm r}{\bf U}\cdot\dot{\bm\Phi}){\rm d}{\bm\Phi}{\rm d}\dot{\bm\Phi}.\label{cond_pf}
\end{align}
This normalization condition eliminates any potential arbitrariness associated with the choice of the argument of delta functions in (\ref{proj_int}). To emphasize that the projection is a conditional expectation value, we will also employ the notation
\begin{equation}
 \hat{P}A(t)\equiv\langle A;{\rm r}(t),{\rm v}(t)\rangle.
\end{equation}
This notation reflects the fact that $\hat{P}$ projects any function $A({\rm r}(t),{\rm v}(t),{\bm\Phi}(t),\dot{\bm\Phi}(t))$ onto the space of functions of (${\rm r}(t),{\rm v}(t)$); by definition, functions $f({\rm r}(t),{\rm v}(t))$ only depending on (${\rm r}(t),{\rm v}(t)$) are left unchanged, i.e. $\hat{P}f({\rm r}(t),{\rm v}(t))=f({\rm r}(t),{\rm v}(t))$.\\

To obtain the key results as directly as possible it is expedient to use the anti-Hermitian Liouville operator $\hat{L}=-\hat{L}^\dag$, which gives the time evolution of a general function $A({\rm r}(t),{\rm v}(t),{\bm\Phi}(t),\dot{\bm\Phi}(t))$ of the crystal configuration through the relation $\frac{\rm d}{{\rm d}t}A\equiv\hat{L}A$. To derive the form of the Liouville operator we simply apply the chain rule-
\begin{align}
\frac{\rm d}{{\rm d}t}A({\rm r}(t),{\rm v}(t),{\bm\Phi}(t),\dot{\bm\Phi}(t))&=
{\rm v}(t)\partial_{\rm r}A+
\dot{\rm v}(t)\partial_{\rm v}A\nonumber\\
&+
\dot{\bm\Phi}\cdot{\bm\nabla}_{\bm\Phi}A+
\ddot{\bm\Phi}\cdot{\bm\nabla}_{\dot{\bm\Phi}}A\nonumber\\
&\equiv\hat{L}A,
\end{align}
which upon substituting in the equations of motion (\ref{DEF_EOM}), (\ref{VIB_EOM}) and requiring the identity to hold for any smooth function results in
\begin{equation}
\hat{L}\equiv-\frac{\partial_{\rm r}V}{\tilde{m}}\partial_{\rm v}+{\rm v}{\partial_{\rm r}}
-\frac{{\bm\nabla}_{\bm\Phi}V}{m}\cdot{\bm\nabla}_{\dot{\bm\Phi}}+\dot{\bm\Phi}\cdot{\bm\nabla}_{\bm\Phi},\label{LIOU}
\end{equation}
where we have introduced the differential operators $\nabla_{{\rm v}}$ and $\nabla_{\dot{\bm\Phi}}$, the velocity space equivalents of $\partial_{\rm r}$ and ${\bm\nabla}_{\bm\Phi}$. Using ${\bm\nabla}_{\dot{\bf X}}$, which acts on the full crystal velocity $\dot{\bf X}$, these operators are defined as in (\ref{P_dx}) and (\ref{VIB_D}), with $\partial_{{\rm v}}\equiv\partial_{\rm r}{\bf U}\cdot{\bm\nabla}_{\dot{\bf X}}$ and $\nabla_{\dot{\bm\Phi}}\equiv\left(\mathbb{I}-\partial_{\rm r}{\bf U}\otimes\partial_{\rm r}{\bf U}/(\partial_{\rm r}{\bf U}\cdot\partial_{\rm r}{\bf U})\right)\cdot{\bm\nabla}_{\dot{\bf X}}$. The formal solution of $\frac{\rm d}{{\rm d}t}A=\hat{L}A$ reads
\begin{equation}
A(t)=\exp\left((t-t')\hat{L}\right)A(t'),
\end{equation}
with the special case $A(t)=\exp(t\hat{L})A(0)$. The exponentiated operator $\exp(t\hat{L})$ may thus be used to evolve any function, or projected function of the system coordinates in time.\\

The first step in our derivation is to partition the dynamical evolution into a subspace  of `fast' and `slow' variables. We have already seen that the projection operator $\hat{P}$ projects any function $A({\rm r}(t),{\rm v}(t),{\bm\Phi}(t),\dot{\bm\Phi}(t))$ onto the space of functions which only take (${\rm r}(t),{\rm v}(t)$) as arguments. The projection operator is idempotent, i.e. $\hat{P}^2=\hat{P}$, so a function $f({\rm r}(t),{\rm v}(t))$ that only depends on (${\rm r}(t),{\rm v}(t)$) is unchanged, i.e. $f=\hat{P}f$. We can also define the complimentary projection operator $\hat{Q}\equiv\mathbb{I}-\hat{P}$, which projects any function $A({\rm r}(t),{\rm v}(t),{\bm\Phi}(t),\dot{\bm\Phi}(t))$ {\it out} of the space of functions which only take (${\rm r}(t),{\rm v}(t)$) as arguments. It is simple to show that $\hat{P}\hat{Q}=\hat{Q}\hat{P}=\hat{P}-\hat{P}=0$. As a result, a function $f({\rm r}(t),{\rm v}(t))$ only depending on (${\rm r}(t),{\rm v}(t)$) vanishes under $\hat{Q}$, i.e. $\hat{Q}f=\hat{Q}\hat{P}f=0$. Just as $\hat{P}$ projected out defect coordinates, $\hat{Q}$ eliminates functions that do not have dependence on the vibrational coordinates. In this way, we can think of the `slow' dynamics being selected by $\hat{P}$ and the `fast' dynamics being selected by $\hat{Q}$.\\
The two projection operators $\hat{P}$ and $\hat{Q}$ give a natural way to partition the defect equation of motion $\tilde{m}\dot{\rm v}=-{\partial_{\rm r}}V({\rm r}(t),{\rm v}(t),{\bm\Phi}(t),\dot{\bm\Phi}(t))=-{\partial_{\rm r}}V(t)$, using $\hat{P}+\hat{Q}=\mathbb{I}$ to write
\begin{equation}
\tilde{m}\dot{\rm v}(t) = -\langle{\partial_{\rm r}} V;r(t)\rangle - \hat{Q}\partial_{\rm r}V(t),
\end{equation}
separating the force into a force expectation $-{\hat P}{\partial_{\rm r}} V(t)=-\langle{\partial_{\rm r}} V;r(t)\rangle$ that depends only on the defect coordinate and a remainder $-\hat{Q}\partial_{\rm r}V(t)=\langle{\partial_{\rm r}} V;r(t)\rangle-\partial_{\rm r}V(t)$ which clearly vanishes under $\hat{P}$ as $\hat{P}\hat{Q}=0$. Using the Liouville operator $\hat{L}$ to extract the time evolution, the defect force then becomes
\begin{equation}
\tilde{m}\dot{\rm v}(t) = -e^{t\hat{L}}\langle{\partial_{\rm r}} V;r(0)\rangle - e^{t\hat{L}}\hat{Q}\partial_{\rm r}V(0),\label{intermed_def_eom}
\end{equation}
where $\exp(t\hat{L})$ is used to evolve the system from its configuration at a time $t=0$. Following a standard method we now use an integral identity\cite{Zwanzig}
\begin{equation}
e^{t\hat{L}}=e^{t\hat{Q}\hat{L}}+\int_0^t{\rm d}se^{(t-s)\hat{L}}\hat{P}\hat{L}e^{s\hat{Q}\hat{L}},\label{dyson}
\end{equation}
that is designed to arrive at the desired result as quickly as possible. This result can be checked by noting the integrand is equal to
$e^{t\hat{L}}\frac{\rm d}{{\rm d}s}(e^{-s\hat{L}}e^{s\hat{Q}\hat{L}})$. Using this identity in (\ref{intermed_def_eom}) gives
\begin{equation}
\tilde{m}\dot{\rm v}(t) = -\langle{\partial_{\rm r}} V;r(t)\rangle+\eta_Q(t)
+\int_0^t{\rm d}s\langle\hat{L}\eta_Q(t);{\rm r}(t-s)\rangle,\label{MT_0}
\end{equation}
where we have defined a `noise' term
\begin{equation}
\eta_Q(t)=e^{t\hat{Q}\hat{L}}\hat{Q}\partial_{\rm r}V(0),
\end{equation}
and a `memory' term $\langle\hat{L}\eta_Q(t);{\rm r}(t-s)\rangle$ which by the definition of the projection operator as a conditional expectation may be written
\begin{align}
\langle\hat{L}\eta_Q(t);{\rm r}(t-s)\rangle&=\int\rho({\rm r}(t-s),{\bm\Phi},\dot{\bm\Phi})\hat{L}\eta_Q(t)\nonumber\\
&\times\delta(\partial_{\rm r}{\bf U}\cdot{\bm\Phi})\delta(\partial_{\rm r}{\bf U}\cdot\dot{\bm\Phi}){\rm d}{\bm\Phi}{\rm d}\dot{\bm\Phi}.\label{MT_1}
\end{align}
The noise term $\eta_Q(t)$ evolves under $e^{t\hat{Q}\hat{L}}$, which can be identified as a `fast' dynamics operator as functions $f(\rm r)$ only of the defect position $\rm r$ are not propagated: $\hat{Q}\hat{L}f({\rm r})={\rm v}\partial_{\rm r}f-\hat{P}{\rm v}\partial_{\rm r}f=0$, meaning that $\exp(t\hat{Q}\hat{L})f({\rm r})=f({\rm r})$.\\

Equation (\ref{MT_0}) is already reminiscent of a non-Markovian Langevin equation with a noise term that vanishes under $\hat{P}$ (meaning it has an expected value of zero) and a memory term that is non-local in time. However, to complete the derivation for our purposes we need to put the integrand of the memory term into a more useful form. Using the anti-Hermitian property $\hat{L}=-\hat{L}^\dag$ of $\hat{L}$ we can act on the distribution function $\rho(t-s)$ in (\ref{MT_1}) instead of $\eta_Q(t)$. We do this in stages for clarity. First, we recognise that
\begin{equation}
\partial_{\rm r}Z({\rm r})=-\beta\langle\partial_{\rm r}V;{\rm r}\rangle Z({\rm r},{\rm v}),
\end{equation}
meaning that $\partial_{{\rm r}}\rho(t)=-\beta\hat{Q}\partial_{\rm r}V$,
which as $\nabla_{{\rm v}}\rho(t)=0$,  gives the identity
\begin{align}
\rho(t-s)\hat{L}&=\hat{L}^\dag\rho(t-s)=-\hat{L}\rho(t-s)\nonumber\\
&= \rho(t-s){\rm v}(t-s)\beta e^{\hat{L}(t-s)}\hat{Q}\partial_{\rm r}V(0).
\end{align}
Defining a second noise term
\begin{equation}
\eta(t) = e^{t\hat{L}}\hat{Q}{\partial_{\rm r}}V(0),
\end{equation}
which evolves under dynamical operator $e^{t\hat{L}}$, unlike $\eta_Q(t)$ that evolves due to $e^{t\hat{Q}\hat{L}}$, we can now re-express (\ref{MT_1}) as
\begin{align}
&\langle\hat{L}\eta_Q(t);{\rm r}(t-s)\rangle= C(s;t){\rm v}(t-s),\\
&C(s;t)=\beta\langle\eta(t-s)\eta_Q(t);{\rm r}(t-s)\rangle,\label{MT_2}
\end{align}
which defines the memory kernel $C(s;t)$ as the time correlation of the defect `noise' forces $\eta(t)$ and $\eta_Q(t)$. In general, the memory kernel retains a dependence on the absolute time $t$ rather than the time difference $s$, as the kernel may depend on the defect position ${\rm r}(t-s)$ used in the expectation average (\ref{MT_2}). However, as we are considering cases where the defect has small migration barriers we expect the average frictional force to vary little with the defect position, i.e. $C(s;t)=C(s)$ an assumption that we have found to be validated in numerical simulations. We can now give a formally exact equation of motion for the defect coordinates as
\begin{equation}
\tilde{m}\ddot{\rm x}(t) = -\langle\partial_{\rm r}V;{\rm r}(t)\rangle+\eta_Q(t)-\int_0^t{\rm d}sC(s){\rm v}(t-s).\label{EXACT_EOM}
\end{equation}
To evaluate the various terms in this equation it would be convenient to replace $\eta_Q(t)$ by $\eta(t)$, thereby eliminating the reduced evolution operator $e^{t\hat{Q}\hat{L}}$ present in $\eta_Q(t)$ in favour of the full evolution operator $e^{t\hat{L}}$ present in $\eta(t)$, as it is $e^{t\hat{L}}$ that evolves the system in MD simulation and thus may be readily calculated. Such an approximation is possible when the defect dynamics occur on a slow time scale $\tau$ over which the vibrational coordinates lose coherence, as in this limit the defect is effectively stationary over the dynamical range of interest, meaning $e^{t\hat{L}}$ and $e^{t\hat{Q}\hat{L}}$ evolve the system in effectively the same manner. In this limit the rate of change of the defect variables ($\hat{P}\hat{L}$) is quantified by the small parameter $\tau^{-1}$, meaning in particular that $\exp({t\hat{Q}\hat{L}})=\exp({t\hat{L}})+O(\tau^{-1})$ and, due to the presence of $\hat{Q}$ in $\eta$ and $\eta_Q$,
\begin{equation}
C(s)=\beta\langle\eta(t-s)\eta(t);{\rm r}(t-s)\rangle+O(\tau^{-3})
\end{equation}
is accurate to order $\tau^{-3}$, see Ref.\cite{Zwanzig}. This is a considerable simplification, because $\eta(t)=\langle\partial_{\rm r}V(t);{\rm r}(t)\rangle
-\partial_{\rm r}V(t)$ is simply the fluctuating part of the defect force, meaning the memory kernel $C(s)$ is simply the autocorrelation of the fluctuating force. As stated above, in this timescale-separated regime we expect $C(s)$ to decay rapidly with $s$. As a result we take the Markovian approximation, where the noise force is replaced by a delta correlated white noise and ${\rm v}(t-s)$ is taken out of the integral in (\ref{EXACT_EOM}). The time integration of $C(s)$ is formally extended to infinity\cite{Zwanzig} though in practice we shall see that a lower limit, usually of order $\tau$, is more appropriate for computational and physical reasons\cite{hijon2010}. With these approximations we finally obtain a Langevin equation for a crystal defect that reads
\begin{equation}
\tilde{m}\dot{\rm v}(t) =-\langle\partial_{\rm r}V;{\rm r}(t)\rangle -{\gamma}{\rm v}(t)+\bar{\eta}(t),\label{eom}
\end{equation}
though it is typical (and usually legitimate\cite{swinburne2013}) in dislocation dynamics to neglect the inertial term $\tilde{m}\dot{\rm v}(t)$, which is valid when the potential energy landscape is slowly varying over the thermal length $\sqrt{{\rm k_BT}/{\tilde{m}\gamma}}$\cite{coffey2004,swinburne2013}. In this limit we finally obtain the overdamped Langevin equation
\begin{equation}
{\rm v}(t)=-\langle\partial_{\rm r}V;{\rm r}(t)\rangle/\gamma+{\eta}(t)/\gamma,\label{eom_od}
\end{equation}
which is identical, after identifying $-\langle\partial_{\rm r}V;{\rm r}(t)\rangle$ with the elastic forces, to (\ref{INTRO_LOD}) given at the start of this paper. In equations (\ref{eom}) and (\ref{eom_od}), $\langle\bar\eta(t)\bar\eta(t')\rangle=2{\rm k_BT}\gamma\delta(t-t')$ is a purely Markovian white noise force\cite{coffey2004} and $\gamma$ is the time integral of the defect force autocorrelation divided by temperature, viz.
\begin{align}
\gamma&=\beta\int_0^\infty C(s){\rm d}s\nonumber\\
&=\beta\int_0^\infty\beta\langle\eta(t-s)\eta(t);{\rm r}(t-s)\rangle.\label{GAMMA_MARK}
\end{align}
Equation (\ref{GAMMA_MARK}) is our main result, a Green-Kubo type relation\cite{reif} that equates the defect force autocorrelation to the defect drag parameter $\gamma$. We have thus achieved a central aim of this paper, to produce a Langevin equation for a crystal defect starting from the classical equation of motion for the 3N atoms of a defective crystal. To see the connection of (\ref{GAMMA_MARK}) to typical Green-Kubo relations, consider the Fokker-Planck equation for a free particle with position $x$ and momentum $p$:
\begin{equation}
\partial_t\rho(x,p,t)=(p/m)\partial_x\rho-\gamma{p}\partial_p\rho+{\rm k_BT}\gamma\partial^2_p\rho.
\end{equation}
The last term has the form of a diffusion operator in momentum space, with a `momentum space diffusivity' ${\rm k_BT}\gamma$\cite{coffey2004}. Therefore, just as the well-known expression $D=\int_0^\infty \langle{\rm v}(s){\rm v}(0)\rangle{\rm d}s$ relates the {\it velocity} autocorrelation $\langle{\rm v}(s){\rm v}(0)\rangle$ to the {\it real} space diffusivity $D$, equation (\ref{GAMMA_MARK}) relates a {\it force} autocorrelation $\int_0^\infty C(s){\rm d}s$ to the {\it momentum} space diffusivity ${\rm k_BT}\gamma$.\\

The real utility of (\ref{GAMMA_MARK}) is that we can calculate the memory kernel directly from simulated trajectories of the stochastic defect force
$-\partial_{\rm r}V(t)=-\partial_{\rm r}{\bf U}\cdot{\bm\nabla}V$. In the present case of small migration barriers the expected force $-\langle\partial_{\rm r} V;{\rm r}(t)\rangle$ will depend only on long range, slowly varying elastic forces which will be uncorrelated with thermal vibrations, contributing a term solely dependent on the defect position. This term will not contribute to $\eta(t)=-\hat{Q}{\partial_{\rm r}}V(t)$ as it will be removed by $\hat{Q}$, meaning that we can invoke a standard ergodicity assumption to write
\begin{align}
C(s)=\beta\lim_{t\to\infty}&\int_0^t\frac{\partial_{\rm r}V(t'+s)\partial_{\rm r}V(t')}{t}{\rm d}t'\nonumber\\
&-\beta\lim_{t\to\infty}\left(\int_0^t\frac{\partial_{\rm r}V(t')}{t}{\rm d}t'\right)^2\label{ts_G}.
\end{align}
Having obtained the Green-Kubo type relation (\ref{GAMMA_MARK}) for the defect drag parameter in terms of the defect force autocorrelation, we now evaluate it in two ways. In section \ref{sec:AG} an analytic derivation of the vibrational expectation values gives the expected autocorrelation (\ref{GAMMA_MARK}) as function of curvatures of the potential energy, which can be evaluated in static calculations in section \ref{sec:FK_I} for the Frenkel-Kontorova model and in section \ref{sec:NI} for realistic crystal defects. This approach gives detailed insight into the form of the interaction between crystal defects and thermal vibrations, explaining the existence of a temperature independent drag coefficient, but numerical evaluation typically requires a full eigenvector decomposition which is expensive for large systems. In section \ref{sec:NI} we present a much more efficient method that directly evaluates the defect force $-\partial_{\rm r}{\bf U}\cdot{\bm\nabla}V$ in molecular dynamics simulations, extracting (\ref{ts_G}) through analysis of the defect force time series.\\
\section{Analytic derivation of $\gamma$}\label{sec:AG}
To derive an expression for $\gamma$ we must remove $\bm\Phi$,$\dot{\bm\Phi}$ by calculating the vibrational expectation values (\ref{PHI_INT}), integrating over the hypersurfaces $\partial_{\rm r}\cdot{\bm\Phi}=0$ and $\partial_{\rm r}\cdot\dot{\bm\Phi}=0$ given in (\ref{cond_x}) and (\ref{cond_v}). In particular, we want to calculate the the defect force autocorrelation $\langle\eta(t-s)\eta(t)\rangle$ for use in (\ref{GAMMA_MARK}), where $\eta(t)=-{\partial_{\rm r}}V(t)+\langle\partial_{\rm r} V\rangle$ is the fluctuating component of the defect force. To do this, we first expand both the potential energy and the defect force in powers of $\bm\Phi$ around ${\bf X}={\mathbf U}({\rm r})$. The expanded potential energy reads
\begin{equation}
V = {V}_0({\rm r})+
\frac{1}{2}{\bm\Phi}\cdot{\bm\nabla}^2_{\bm\Phi}V_0\cdot{\bm\Phi}
+
\frac{1}{3!}{\bm\Phi}\cdot{\bm\nabla}^3_{\bm\Phi}V_0\cdot{\bm\Phi}\cdot{\bm\Phi}+...
,\label{cryv}
\end{equation}
where the subscript $V_0$ indicates that all the partial derivatives are calculated at ${\bf X}={\bf U}({\rm r})$. The first vibrational derivative ${\bm\nabla}_{\bm\Phi}V_0=0$ as by definition $\langle\bm\Phi\rangle=0$ at zero temperature.\\

In order to perform an integration over $\bm\Phi$,$\dot{\bm\Phi}$ we must parametrise the hypersurface of allowed values, namely all directions orthogonal to $\partial_{\rm r}{\bf U}$. To do this we define a basis set of $3N-1$ orthonormal vectors $\{{\bf e}_k(\rm r)\}_{k=1}^{3N-1}$ which are all orthogonal to $\partial_{\rm r}{\bf U}$, i.e. $\partial_{\rm r}{\bf U}\cdot{\bf e}_k({\rm r})=0$ $\forall$ $k$. In order for this condition to hold for all values of the defect position the $\{{\bf e}_k\}$ will vary with $\rm r$, though as we perform the integrals for a given value of the defect position this will not affect the result of our integration. With a set of $3N-1$ vibrational amplitudes $\{\phi_k\}_{k=1}^{3N-1}$ and momenta $\{p_k\}_{k=1}^{3N-1}$ we are free to express $\bm\Phi$ and $\dot{\bm\Phi}$ as
\begin{equation}
{\bm\Phi}(t)=\sum_{k=1}^{3N-1}\phi_k(t){\bf e}_k({\rm r}),\, m\dot{\bm\Phi}(t)=\sum_{k=1}^{3N-1}p_k(t){\bf e}_k({\rm r}),\label{PHI_phi}
\end{equation}
which explicitly shows the dependence of the vibrational co-ordinates on $\rm r$. We can use this parametrization to evaluate (\ref{proj_int}) through a change of variables in the integration. As the $\{{\bf e}_k\}$ are by definition orthonormal the Jacobian in (\ref{proj_int}) is simply unity for $\bm\Phi$ and $1/m^{3N-1}$ for $\dot{\bm\Phi}$, giving
\begin{align}
\hat{P}A(t)&\equiv\int A({\rm r},{\rm v},\{\phi_k,p_k\})
\rho\prod_k^{3N-1}{\rm d}\phi_k({\rm d}p_k/m),\\\label{PHI_INT}
&\equiv\langle A;{\rm r}(t),{\rm v}(t)\rangle.
\end{align}
It is convenient to also define the differential operator
\begin{equation}
\nabla_{\phi_k}\equiv{\bf e}_k\cdot{\bm\nabla},
\end{equation}
which, as the $\{{\bf e}_k\}$ span all directions orthogonal to $\partial_{\rm r}{\bf U}$ gives the identity
\begin{equation}
\mathbb{I}-\partial_{\rm r}{\bf U}\otimes\partial_{\rm r}{\bf U}/(\partial_{\rm r}{\bf U}\cdot\partial_{\rm r}{\bf U})=\sum_{k=1}^{3N-1}{\bf e}_k\otimes{\bf e}_k.\label{vib_ev_sum}
\end{equation}
As the differential operator ${\bm\nabla}_{\bm\Phi}$ is by construction orthogonal to $\partial_{\rm r}{\bf U}$ (see equation (\ref{VIB_D})) we can therefore expand ${\bm\nabla}_{\bm\Phi}$ as
\begin{equation}
{\bm\nabla}_{\bm\Phi}=\sum_{k=1}^{3N-1}{\bf e}_k\nabla_{\phi_k}.
\end{equation}
In the remaining manipulations we will employ the Einstein summation convention\cite{Peskin}, where repeated indices are taken to be summed from $1$ to $3N-1$. For example, the expansion of differential operator ${\bm\nabla}_{\bm\Phi}$ can now be written
\begin{equation}
{\bm\nabla}_{\bm\Phi}={\bf e}_k\nabla_{\phi_k}.
\end{equation}
With these identities we can now rewrite the expanded potential energy, using the Einstein summation convention, as
\begin{equation}
V = {V}_0({\rm r})+
\frac{\phi_q\phi_r}{2}{\nabla}^2_{\phi_q\phi_r}V_0
+
\frac{\phi_q\phi_r\phi_s}{3!}{\nabla}^3_{\phi_q\phi_r\phi_s}V_0+..,\label{cryv_2}
\end{equation}
where again all the partial derivatives are evaluated at ${\bf X}={\bf U}(\rm r)$. This last expansion immediately suggests an expedient choice for the basis vectors $\{{\bf e}_k\}$: the eigenbasis of the matrix of second derivatives in (\ref{cryv}), the `$\Phi$-Hessian'
\begin{equation}
{\bm\nabla}^2_{\bm\Phi}V\equiv
\left(\mathbb{I}-\frac{\partial_{\rm r}{\bf U}\otimes\partial_{\rm r}{\bf U}}{
\partial_{\rm r}{\bf U}\cdot\partial_{\rm r}{\bf U}
}\right)
\cdot{\bm\nabla}^2V\cdot
\left(\mathbb{I}-\frac{\partial_{\rm r}{\bf U}\otimes\partial_{\rm r}{\bf U}}{\partial_{\rm r}{\bf U}\cdot\partial_{\rm r}{\bf U}}\right).\label{VIB_HESS}
\end{equation}
With knowledge of $V({\bf X})$ and $\partial_{\rm r}{\bf U}$ this real, symmetric matrix is readily constructed and can be fully diagonalized to obtain $3N$ orthonormal eigenvectors and eigenvalues. The $\Phi$-Hessian ${\bm\nabla}^2_{\bm\Phi}V$ is distinct from the full Hessian ${\bm\nabla}^2V$ as the vector of defect motion $\partial_{\rm r}{\bf U}$ is by construction an eigenvector with an eigenvalue of identically zero, meaning that the $3N-1$ remaining eigenvectors form an orthonormal set $\{{\bf e}_k(\rm r)\}$, orthogonal to $\partial_{\rm r}{\bf U}$, with `eigenfrequencies' $\{\omega_k\}$ such that
\begin{equation}
{\bm\nabla}^2_{\bm\Phi}V \cdot {\bf e}_k=({\nabla}^2_{\phi_k}V_0){\bf e}_k=m\omega^2_k{\bf e}_k.
\end{equation}
This choice of $\{{\bf e}_k\}$ as the eigenbasis of ${\bm\nabla}^2_{\bm\Phi}V$ means that ${\nabla}^2_{\phi_q\phi_r}V_0=\delta_{qr}m\omega^2_q$ is diagonal. If there is no external potential the crystal will have global translational symmetry, meaning three eigenvectors will always be closely related to a global rigid translation with extremely small eigenvalues; we will omit these when calculating vibrational expectation values. After this choosing the vibrational basis vectors to be the eigenvectors of the $\bm\Phi$-Hessian ${\bm\nabla}^2_{\bm\Phi}V$ the expanded potential energy takes the simpler form
\begin{equation}
V = {V}_0({\rm r})+\frac{\phi^2_q}{2}m\omega^2_q+\frac{\phi_q\phi_r\phi_s}{3!}{\nabla}^3_{\phi_q\phi_r\phi_s}V_0+..,\label{cryv_ind}
\end{equation}
whilst the defect force expansion reads
\begin{equation}
{\partial_{\rm r}}{V} ={\partial_{\rm r}}{V}_0
+
\phi_q\partial_{\rm r}{\nabla}_{\phi_q}V_0
+
\frac{\phi_q\phi_r}{2}\partial_{\rm r}{\nabla}_{\phi_q}\nabla_{\phi_r}V_0
+..,\label{cryf}
\end{equation}
where in both cases the Einstein summation convention is used. Practically, the mixed derivatives in (\ref{cryf}) are evaluated by contracting the full tensorial derivative ${\bm\nabla}^nV$, evaluated at ${\bf X}={\bf U}(\rm r)$, once with $\partial_{\rm r}{\bf U}$ and $n-1$ times with the $\{{\bf e}_k\}$. It is important to note that although ${\nabla}_{\phi_q}V=0$, there is no requirement that the {\it mixed} derivatives vanish; moreover, these terms give the coupling between the defect and thermal vibrations. To proceed, we now truncate $V$ to quadratic vibrations in the Gibbs distribution (\ref{FE}). Our choice $\{{\bf e}_k\}$ means that the Gibbs distribution is a product of Gaussians, namely
\begin{equation}
\rho({\rm r},\{\phi_k,p_k\}) =  e^{-\beta\left[V_0({\rm r})+m\sum_q\omega^2_q{\phi^2_q}/{2}+\sum_q{p^2_q}/{2m}\right]}/Z({\rm r}),\label{FE_phi}
\end{equation}
meaning expectation values may be calculated using standard identities such as Wick's theorem\cite{Peskin}. This truncation clearly neglects any thermal expansion arising from the purely vibrational anharmonicities $\bm\nabla^{n\geq3}_{\bm\Phi}V$. In a previous publication\cite{swinburne2014} we systematically included these terms to produce an expression for $\gamma$ up to linear order in temperature. It was shown that our main result, the anomalous temperature independent drag coefficient $\gamma_0$ in equation (\ref{full_gamma}) below, is unaffected by these additional terms.\\

To analytically evaluate the defect force autocorrelation and hence the memory kernel $C(s)$, we evolve the vibrational coordinates from a fixed ${\rm r}$. This is justified by the time-scale separation, as $C(s)$ is expected to decay to incoherence before $\rm r$ changes appreciably, an assumption we will test in MD simulation. To do this we truncate the vibrational force to linear order in (\ref{VIB_EOM}), differentiate (\ref{PHI_phi}) and invoke the timescale separation to neglect terms of order ${\rm v}$, obtaining an {\it approximate} vibrational equation of motion of
\begin{equation}
\ddot{\phi}_q=-m\omega^2_q\phi_q.\label{PHI_EOM}
\end{equation}
We emphasize that this equation of motion only holds over short time periods where the defect co-ordinate $\rm r$ is effectively stationary and the defect force autocorrelation $C(s)$ is expected to decay to incoherence, an assumption we will test explicitly in section \ref{sec:NI}. Over longer periods of time $\rm r$, and hence the vibrational basis $\{{\bf e}_k(\rm r)\}$, will change; nevertheless, on the short timescale of present interest (\ref{PHI_EOM}) may be solved with initial conditions $\langle{\phi}_q(0){\phi}_r(0)\rangle=({\rm k_BT}/{m\omega^2_q})\delta_{qr}$, $\langle\dot{\phi}_q(0){\phi}_r(0)\rangle=0$, allowing us to explicitly evaluate time dependent expectation values of $\{\phi_k\}$ in terms of the two point correlation functions\cite{Peskin}
\begin{equation}
\langle\phi_q(t)\phi_r(0)\rangle=\frac{{\rm k_BT}}{m\omega^2_q}\delta_{qr}\cos(\omega_qt).\label{CF}
\end{equation}
As appropriate for non-conservative dynamics, the correlation function is evaluated using only initial conditions and consequently is closely related to the {\it retarded} Green's function\cite{galley2013}
\begin{equation}
{\rm G}_q(t)=\Theta(t)\beta\langle\phi_q(t)\phi_q(0)\rangle=\frac{\Theta(t)}{m\omega^2_q}\cos(\omega_qt),
\end{equation}
where $\Theta(t)$ is the Heaviside step function. Finally, using Wick's theorem\cite{Peskin} to simplify high order expectation values $\langle\phi_q\phi_r\phi_s\rangle$, $\langle\phi_q\phi_r\phi_s\phi_t\rangle$ in the force autocorrelation, we perform the elementary Gaussian integrations to give our main result
\begin{align}
{\gamma} &= \int_0^\infty (\partial_{\rm r}{\nabla}_{\phi_q}V)^2{\rm G}_q(t){\rm d}t\nonumber\\
&+\frac{k_{\rm B}T}{2}\int_0^\infty\partial_{\rm r}{\nabla}^2_{\phi_q\phi_r}V{\rm G}_r(t){\rm G}_s(t)\partial_{\rm r}{\nabla}^2_{\phi_s\phi_q}V{\rm d}t\nonumber\\
&+\frac{k_{\rm B}T}{2}\int_0^\infty\partial_{\rm r}{\nabla}_{\phi_q}V{\rm G}_q(t){\rm G}_r(t)\partial_{\rm r}{\nabla}^3_{\phi_q\phi_r\phi_r}V\,{\rm d}t
\label{full_gamma}
\end{align}
We see that the friction coefficient takes the form $\gamma = {\gamma}_0+{\rm k_B T}{\gamma}_{\rm w}$, with a {\it temperature independent} component
\begin{equation}
\gamma_0 = \int_0^\infty \frac{(\partial_{\rm r}{\nabla}_{\phi_q}V)^2}{m\omega^2_q}\cos(\omega_qt){\rm d}t.\label{full_gamma_0}
\end{equation}
The temperature independent drag coefficient $\gamma_0$ couples the defect to thermal vibrations via a mixed {\it quadratic} derivative
\begin{equation}
\partial_{\rm r}{\nabla}_{\phi_k}V=\partial_{\rm r}{\bf U}\cdot{\bm\nabla}^2V\cdot{\bf e}_k,\label{gamma_0_coupling}
\end{equation}
where ${\bm\nabla}^2V$ is the full Hessian. This is the first time, to our knowledge, that a theoretical derivation has produced an expression for the widely observed temperature independent drag force acting on a crystal defect.\\

The existence of this coupling means that vibrational displacements $\bm\Phi$ that are orthogonal to defect motion (as $\partial_{\rm r}{\bf U}\cdot\bm\Phi=0$) can still induce a force on the defect (through $-\partial_{\rm r}{\bf U}\cdot{\bm\nabla}^2V\cdot\bm\Phi$) to {\it linear} order in the vibrational displacements. As a result any theory that assumes that thermal vibrations are harmonic oscillators perturbed by anharmonic terms will never capture $\gamma_0$ as it is assumed {\it a priori} not to exist. In our approach, although we approximate the dynamics of $\bm\Phi$ by (\ref{DEF_EOM}) over short timescales, our general derivation made no assumption on the form of the crystal potential energy and thus allows any form of coupling to exist.\\

The temperature dependent drag coefficient ${\rm k_BT}\bm\gamma_{\rm w}$ couples through the mixed cubic and quartic derivatives $\partial_{\rm r}{\nabla}^{2}_{\phi}V$ and $\partial_{\rm r}{\nabla}^{3}_{\phi}V$. As typical scattering theories only include cubic anharmonic\cite{brailsford1970,brailsford1972,alshits1975} only the second term in (\ref{full_gamma}) is directly comparable. In a rough approximation, if we consider the Hessian as approximately ${\bm\nabla}^2V\sim \sum m\omega^2$ cubic anharmonicities become ${\bm\nabla}^3V\sim\sum 2m\omega(\nabla\omega)$, meaning the cubic coupling term is approximately $({\bm\nabla}^3V)^2/m\omega^2\sim(\sum(\nabla\omega)/\omega)^2$, roughly comparable to square of the Gr\"{u}neisen parameter\cite{Ashcroft} $\gamma_{\rm G}=\sum\frac{\partial\omega}{\partial V}/\omega$, which measures the change of vibrational frequencies with hydrostatic pressure. As a result we conclude that the temperature dependent frictional coupling is of order ${\rm k_BT}\gamma_{\rm w}\sim{\rm k_BT}(\gamma_{\rm G})^2$ in agreement with other approximate analyses\cite{brailsford1972}. The precise nature of the anharmonic coupling will be the subject of a future publication. However, in the remainder of our analysis we concentrate on the nature of the anomalous temperature independent $\gamma_0$.
\section{A special test case: kinks in the Frenkel-Kontorova model}\label{sec:FK_I}
We have seen in section \ref{sec:PS_INTRO} how phonon scattering theories forbid a temperature independent drag parameter, $\gamma_0=0$, as thermal vibrations are taken to be the phonon modes of the perfect lattice with, before the inclusion of anharmonic vibrations, a conserved energy and momentum. To better understand how the assumptions of phonon scattering theory contrast with the present approach, we investigate a rare system which supports localised deformations but also admits, in the continuum limit, a full analytic evaluation of all the terms ($\partial_{\rm r}{\bf U},\{{\bf e}_q\},{\bm\nabla}^2_{\bm\Phi}V$, etc.) used in our analysis. This will highlight the precise effect of canonical thermal vibrations on the motion of crystal defects.\\

The Frenkel-Kontorova model\cite{Braun2004,Kosevich2006} is built from a set of $N$ harmonically coupled nodes with one-dimensional positions ${\bf X}=(X_0,X_2,...,X_{N-1})\in\mathbb{R}^N$, sitting in a sinusoidal `lattice' potential of period $b$. The potential energy reads\cite{Kosevich2006,swinburne2013b}
\begin{equation}
V_{\rm FK}({\bf X})=\sum_ia\frac{\kappa}{2}\frac{({\rm X}_{i+1}-{\rm X}_{i}-a)^2}{a^2}+a\nu\sin^2\left(\pi\frac{{\rm X}_i}{b}\right),\label{E_FK}
\end{equation}
where $a$ is the equilibrium spacing of the harmonic coupling and $\kappa$, $\nu$ are energy densities that control the relative strength of the coupling and `lattice' potential. The kinetic energy is simply $\sum_ia(\mu/2)(\dot{X}_i)^2$, where $a\mu$ is the node mass and the system is typically completed with periodic boundary conditions ${\rm X}_N={\rm X}_0+Na$. Dynamics are classical, i.e. produced by
\begin{equation}
a\mu\ddot{\rm X}_i=-\nabla_{\rm X_i}V_{\rm FK}({\bf X}).
\end{equation}
The continuum limit takes the discrete chain of nodes $\{{\rm X}_n\}$ to an elastic line ${\rm X}(x)$, such that ${\rm X}_n\to na+{\rm X}(na)$. As $a\to0$, $({\rm X}_{i+1}-{\rm X}_{i}-a)/a\to\nabla{\rm X}(x)$, $Na\to L\to\infty$ and $\sum_ia\to\int{\rm d}x$. As with all such limits, directions in a vector space now become functions in a Hilbert space\cite{Peskin}, with inner products becoming integrals. In this limit we obtain the so called sine-Gordon model
\begin{equation}
V[{\rm X}]=\int_{0}^{L}{\rm d}x\frac{\kappa}{2}(\nabla{\rm X}(x))^2+\nu\sin^2\left(\pi\frac{{\rm X}(x)}{b}\right),\label{E_SG}
\end{equation}
with a kinetic energy $\int_{0}^{L}{\rm d}x(\mu/2)(\dot{{\rm X}}(x))^2$. It is well known that with boundary conditions ${\rm X}(x+L)={\rm X}(x)+a$ the sine-Gordon model support solitons of the form ${\rm X}(x)={\rm U}(x,{\rm r})$, where the defect configuration ${\rm U}(x,{\rm r})$ reads\cite{swinburne2013}
\begin{equation}
{\rm U}(x,{\rm r})=\frac{a}{2}+\frac{a}{\pi}\arctan\left(\sinh\left(\frac{x-{\rm r}}{w}\right)\right).\label{SOLIT}
\end{equation}
The soliton width $w$ is given by $w=(b/2\pi)\sqrt{2\kappa/\nu})$ (we assume $L\gg w$) and the soliton energy is $E=(2b/\pi)\sqrt{2\kappa\nu}$, which is clearly independent of solition position, i.e. the migration barrier is identically zero. It is a simple matter to give the direction of defect motion (now in Hilbert space) as
\begin{equation}
\partial_{\rm r}{\rm U}(x,{\rm r})=-\frac{a}{\pi{w}}{\rm sech}\left(\frac{x-{\rm r}}{w}\right),\label{d_solit}
\end{equation}
giving an effective mass of
\begin{equation}
\tilde{m}=\mu\int_{0}^{L}\left(\partial_{\rm r}{\rm U}(x,{\rm r})\right)^2{\rm d}x=\frac{2a^2\mu}{\pi^2w}.
\end{equation}
The sine-Gordon model is an integrable system, which is particularly remarkable in that one can derive exact analytical forms for the set ${\rm e}_q(x,{\rm r})$ of the vibrational modes orthogonal to $\partial_{\rm r}{\rm U}$ which remains valid {\it for all values of} ${\rm r}$. The vibrational modes are given by\cite{Kosevich2006}
\begin{equation}
{\rm e}_q(x,{\rm r}) = \left(\frac{\pi{q}w}{L}+i\tanh\left(\frac{x-{\rm r}}{w}\right)\right)\exp\left({i\frac{\pi{qx}}{L}-i\omega_qt}\right),
\end{equation}
where $\omega_q=\sqrt{\kappa/\mu}\sqrt{1+(\pi{q}w/L)^2}$, meaning that thermal vibrations can be included as
\begin{align}
\Phi(x,{\rm r})&=\int_{0}^L{\rm d}q\phi_q{\rm e}_q(x,{\rm r}),\label{SG_PHI}\\
m\dot{\Phi}(x,{\rm r})&=\int_{0}^L{\rm d}qp_q{\rm e}_q(x,{\rm r}),
\end{align}
thereby giving analytic expressions for all the quantities used in the theory developed in section \ref{sec:AG}. It is simple to show that the orthogonality conditions (\ref{cond_x}), (\ref{cond_v}) are satisfied as\cite{Kosevich2006}
\begin{equation}
\int_{0}^{L}{\rm d}x\partial_{\rm r}{\rm U}(x,{\rm r}){\rm e}_q(x,{\rm r})=0.\label{SG_ORTH}
\end{equation}
However, $\partial_{\rm r}{\rm U}$ and the $\{{\rm e}_q(x,{\rm r})\}$ are also eigenmodes of the {\it full} Hessian $\nabla^2V(x,x')=\delta(x-x')\nabla^2V(x)$, where
\begin{equation}
\nabla^2V(x)=\kappa\nabla^2+\frac{\pi^2\nu}{a^2}\frac{\sinh^2\left((x-{\rm r})/{w}\right)-1}{\cosh^2\left((x-{\rm r})/{w}\right)}.
\end{equation}
$\partial_{\rm r}{\rm U}$ has an eigenvalue of identically zero
\begin{align}
&\int_{0}^L\nabla^2V(x,x')\partial_{\rm r}{\rm U}(x',{\rm r}){\rm d}x'=\nabla^2V(x)\partial_{\rm r}{\rm U}(x,{\rm r}),\nonumber\\\label{SG_DU_EV}
&=\frac{2\sinh^2(\frac{x-{\rm r}}{w})-\cosh^2(\frac{x-{\rm r}}{w})}{\cosh^3(\frac{x-{\rm r}}{w})}\left(\frac{a}{\pi w^3}-\frac{a}{\pi w^3}\right)=0,
\end{align}
whilst the vibrational modes ${\rm e}_q(x,{\rm r})$ have eigenvalues of $\mu\omega^2_q$, namely
\begin{align}
\int_{0}^L\nabla^2V(x,x'){\rm e}_q(x',{\rm r}){\rm d}x'&=\nabla^2V(x){\rm e}_q(x,{\rm r})\nonumber\\
&=\mu\omega^2_q{\rm e}_q(x,{\rm r}).
\end{align}
We thus see that the sine-Gordon model is a very special case of the general derivation of section \ref{sec:AG}, where we found that thermal vibrations $\bm\Phi$ could be decomposed in the eigenmodes $\{{\rm e}_q({\rm r})\}$ of the $\bm\Phi$-Hessian (\ref{VIB_HESS}) evaluated at a given value of $\rm r$. In the special case of the sine-Gordon model, whilst it is still true that the thermal vibrations $\Phi(x,{\rm r})$ can be decomposed into the eigenmodes ${\rm e}_q(x,{\rm r})$ of the $\Phi$-Hessian $\nabla_{\rm\Phi}^2V(x,x')$, in this case the $\Phi$-Hessian $\nabla_{\rm\Phi}^2V(x,x')$ is equal to the full Hessian $\nabla^2V(x,x')$. To see this, we first construct the Hilbert space `matrix' $Q(x,x',{\rm r})$ of all directions orthogonal to $\partial_{\rm r}{\rm U}$, namely
\begin{equation}
Q(x,x',{\rm r})=\delta(x-x')-\frac{\partial_{\rm r}{\rm U}(x,{\rm r})\partial_{\rm r}{\rm U}(x',{\rm r})}{\int_{0}^{L}\left(\partial_{\rm r}{\rm U}(x,{\rm r})\right)^2{\rm d}x},
\end{equation}
meaning that the $\Phi$-Hessian $\nabla_{\rm\Phi}^2V(x,x')$ reads
\begin{equation}
\nabla^2_{\rm\Phi}V(x,x')=\int_{0}^{L}\int_{0}^{L}Q(x,y,{\rm r})\nabla^2V(y,z) Q(z,x',{\rm r}){\rm d}y{\rm d}z,\label{SG_VIB_HESS}
\end{equation}
which is the continuum equivalent of the discrete $\Phi$-Hessian (\ref{VIB_HESS}). However, because $\partial_{\rm r}{\rm U}$ is an eigenmode of the {\it full} Hessian with eigenvalue zero, as shown in (\ref{SG_DU_EV}), we have that
\begin{equation}
\int_{0}^{L}Q(x,y,{\rm r})\nabla^2V(y,z){\rm d}y=\nabla^2V(x,z),
\end{equation}
and similarly for the other inner product, giving the final result that
\begin{equation}
\nabla^2_{\rm\Phi}V(x,x')=\delta(x-x')\nabla^2V(x)=\nabla^2V(x,x'),
\end{equation}
showing the equivalence of $\nabla^2_{\rm\Phi}V(x,x')$ and $\nabla^2V(x,x')$ for this system. In this rare case of an integrable system we thus have a convergence of the approach developed here and the assumptions of phonon scattering theory, indeed we can use the correct expression for the vibrational modes (rather than plane waves of scattering theory) and treat the vibrational momentum and energy as conserved quantities (rather than, as in our approach, simply an expression of the crystal configuration contingent with a given $\rm r$). As a result, in this case the eigenmodes ${\rm e}_q(x,{\rm r})$ of the full Hessian {\it are} orthogonal to the defect motion mode  $\partial_{\rm r}{\rm U}$, as shown in equation (\ref{SG_ORTH}). From our analytical result (\ref{full_gamma_0}) we have seen that the temperature-independent drag parameter $\gamma_0$, is a sum over all the modes of the $\Phi$-Hessian $\nabla_{\rm\Phi}^2V$ weighted by a coupling term (\ref{gamma_0_coupling}) $\partial_{\rm r}{\bf U}\cdot{\bm\nabla}^2V\cdot{\bf e}_k$. In the Sine-Gordon model, it is simple to show that this coupling vanishes, viz.
\begin{align}
\partial_{\rm r}{\bf U}\cdot{\bm\nabla}^2V\cdot{\bf e}_k &\to \int_0^L{\rm d}x
\partial_{\rm r}{\rm U}(x,{\rm r})\nabla^2_xV{\rm e}_q(x,{\rm r}),\\
 &=\omega^2_q\int{\rm d}x\partial_{\rm r}{\rm U}(x,{\rm r}){\rm e}_q(x,{\rm r})=0,\label{SG_G0}
\end{align}
which in turn means that the temperature independent drag parameter $\gamma_0=0$ vanishes. This result is a direct consequence of the integrability of the sine-Gordon model, a rare property that allows one to identify a general eigenmode expansion encompassing the direction of defect motion $\partial_{\rm r}{\rm U}$ and all other orthogonal vibrational modes ${\rm e}_q(x,{\rm r})$, valid for all values of the defect position. As the vibrational modes are now modes of the {\it full} Hessian rather than just the $\Phi$-Hessian, the amplitude $\phi_q$ of each vibrational modes can be considered as an isolated harmonic oscillator, independent of any defect before the inclusion of anharmonic vibrational terms, with a conserved energy and momentum.\\

Phonon scattering calculations also assume that phonons are harmonic oscillators with a conserved energy and momentum, interacting with defects only after the inclusion of anharmonic vibrational terms. But we have just seen that for these assumptions to hold in our general approach we are forced to consider only integrable systems, rare non-linear models that typically only exist in one dimension\cite{Braun2004}. It is therefore very interesting that only in this somewhat abstract limit does our general approach come into agreement with the central tenet of any phonon scattering calculation. This highlights the `artificial integrability' that scattering calculations must impose on a problem in order to have a well defined phase space for phonon scattering. In general we do not expect these assumptions to hold and thus we anticipate $\gamma_0\neq0$ to be a general feature of defects in any realistic crystalline system.\\

These considerations promote the Frenkel-Kontorova (\ref{E_FK}) model as an interesting test case to calculate $\gamma_0$, as by better approximating the continuum limit we reach the Sine-Gordon model, where by equation (\ref{SG_G0}) $\gamma_0$ should vanish. To this end we have performed static calculations on the Frenkel-Kontorova model (\ref{E_FK}), which in appropriate regions of parameter space also supports soliton-like `kink' deformations under boundary conditions ${\rm X}_{n+N}={\rm X}_{n}+b$. To investigate the continuum limit we scale the line tension $\kappa$ and lattice barrier $\nu$ in inverse proportion, i.e. $(\kappa,\nu)\to(\alpha\kappa,\nu/\alpha)$ such that the kink energy $\sim\sqrt{\kappa\nu}$ is constant and the kink width $\sim\sqrt{\kappa/\nu}\to\alpha\sqrt{\kappa/\nu}$ increases; in this way the deformations of the chain become smoother and better approximate the continuum field of the sine-Gordon model, whilst as the kink energy is roughly constant we can compare results taken for different values of the kink width.\\

It is well known\cite{Braun2004,swinburne2013} that for kink widths greater than a few node spacings (i.e. $w/a\gtrsim2$ the system state is very well approximated by the discretised soliton kink solution ${\rm X}_n={\rm U}_n({\rm r})=na+{\rm U}(na-{\rm r})$, where ${\rm U}(na-{\rm r})$ is given by (\ref{SOLIT}). However, when we substitute this solution into the form of the total energy (\ref{E_FK}) the discrete summation retains a dependence on kink position, giving a system energy of approximately\cite{swinburne2013}
\begin{equation}
E({\rm r})=(2b/\pi)\sqrt{2\kappa\nu}+\frac{2\pi{w^2}\nu}{a}\exp(-\pi{w/a})\sin^2(\pi{\rm r}/a),\label{FK_KINK}
\end{equation}
being the continuum result $(2b/\pi)\sqrt{2\kappa\nu}$ plus a periodic term, of period $a$, that decays exponentially fast with the kink width $w$. The presence of this periodic migration barrier is an accepted signature of discreteness effects in crystalline systems, with the exponential decay with defect width famously derived by Peierls\cite{PeierlsOld} to explain the low critical shear stress of dislocations.\\

The kink migration barrier decays rapidly with increasing kink width as a broader kink profile causes the summation in the total energy (\ref{E_FK}) to better approximate the integral (\ref{E_SG}) of the continuum limit. As this integral has no dependence on kink position the kink migration barrier vanishes. It is interesting to ask whether we expect a similar dependence for the temperature independent drag parameter $\gamma_0$.\\

Whilst the kink migration barrier arises from the kink formation energy being a function of defect position in a static system, the expression (\ref{full_gamma_0}) for $\gamma_0$, in contrast, is a sum across all the fluctuations in the system, weighted by a coupling term $(\partial_{\rm r}{\bf U}\cdot{\bm\nabla}^2V\cdot{\bf e}_q)^2$.

In the continuum limit of the Sine-Gordon model we have seen (equation (\ref{SG_G0})) that this coupling term vanishes, as the $\{{\rm e}_q(x,{\rm r})\}$ are eigenmodes of the full Hessian $\nabla^2V(x)\delta(x-x')$, namely that $\int_0^L\nabla^2V(x,x'){\rm e}_q(x',{\rm r})=\mu\omega^2_q{\rm e}_q(x,{\rm r})$. In this case the discrete summation that approximates $\int_0^L\nabla^2V(x,x'){\rm e}_q(x',{\rm r})$ is the inner product ${\bm\nabla}^2V\cdot{\bf e}_q$, where the summation is over all elements $[({\rm e}_q)_1,({\rm e}_q)_2,...({\rm e}_q)_N]$ of the vector ${\bf e}_q$. However, unlike the kink profile, the fluctuations ${\bf e}_q$ in a discrete system have no requirement to vary slowly between lattice sites, regardless of the kink width or any defects the same system. As a result we cannot say that an increasing kink width suppresses discreteness effects in the fluctuations, meaning we expect $\gamma_0$ to decay much more slowly than the migration barrier as we attempt to approximate the continuum limit. \\

We emphasize that this is a general feature; even if discreteness effects are negligible for the migration of stable defect structures that possess a broad displacement field, the {\it fluctuations} around a crystal defect of any size will always be sensitive to discreteness as they have no requirement vary slowly across lattice sites and thus approximate a continuum solution.\\

To test these conclusions we have evaluated $\gamma_0$ for a kink in the Frenkel-Kontorova model, employing the techniques detailed in section \ref{sec:NI} below. We construct and fully diagonalize ${\bm\nabla}^2_{\bm\Phi}V$ to obtain $\{{\bf e}_q,\omega_q\}$ and ${\bf U}({\rm r})$, allowing us to evaluate the integrand in (\ref{full_gamma_0}), which is then integrated with respect to time (see below) to produce $\gamma_0$. In agreement with our earlier assumption that the frictional force is essentially independent of defect position, the actual value of $\gamma_0$ was found to vary very little with the precise position of the defect ${\rm r}$.\\

\begin{figure}
\centering
\includegraphics[width=0.48\textwidth]{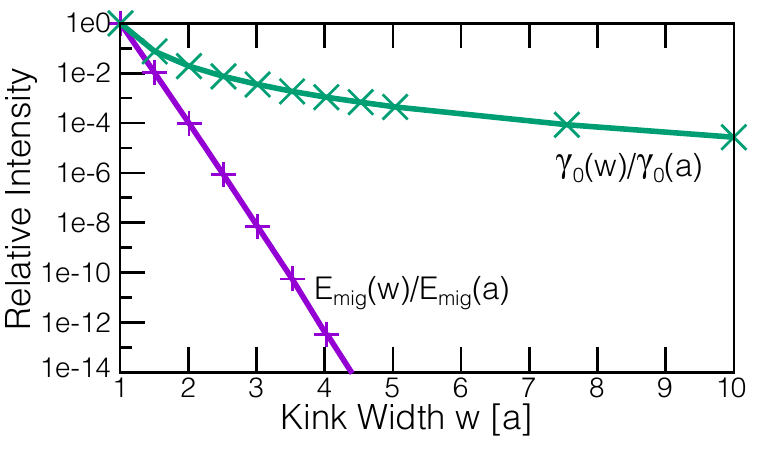}
\caption{Migration barrier $E_{\rm mig}$ and temperature independent friction coefficient $\gamma_0$ as a function of kink width $w$, relative to their values at $w=a$. We see the migration barrier decays exponentially, whilst the temperature independent friction coefficient remains almost unchanged even for kink widths much larger than the lattice parameter $a$. This highlights the fact that $\gamma_0$ is a quantity owing its existence to the discreteness of the system, which is present even when the energy cost of discreteness is low. (Colour Online)\label{FKG}}
\end{figure}

The results of these calculations are shown in Figure \ref{FKG}. We see that whilst the temperature independent drag coefficient does indeed eventually vanish as the continuum limit is reached (with increasing kink width) it does so at a much slower rate than the kink migration barrier, which rapidly becomes indistinguishable from zero. We therefore conclude that a temperature independent friction parameter is robust, arising due to the coupling of a localised defect to {\it fluctuations} in a {\it discrete} system. These fluctuations give an important discreteness effect even in finely interpolated approximations to integrable systems, where they are expected to vanish, and typical {\it static} signatures of discreteness such as migration barriers are indistinguishable from zero. This demonstrates that a temperature independent drag coefficient $\gamma_0$ arises even when migration barriers vanish, invalidating a proposed `radiative damping' mechanism that argues $\gamma_0$ arises due to a defect radiating phonons when traversing migration barriers. As $\gamma_0$ is present even in this idealised system it is reasonable to expect this phenomenon to arise for all crystal defects, which we investigate in the next section, applying the developed theory to realistic crystalline systems.
\section{Numerical evaluation of $\gamma$}\label{sec:NI}
\begin{figure}
\includegraphics[width=0.48\textwidth]{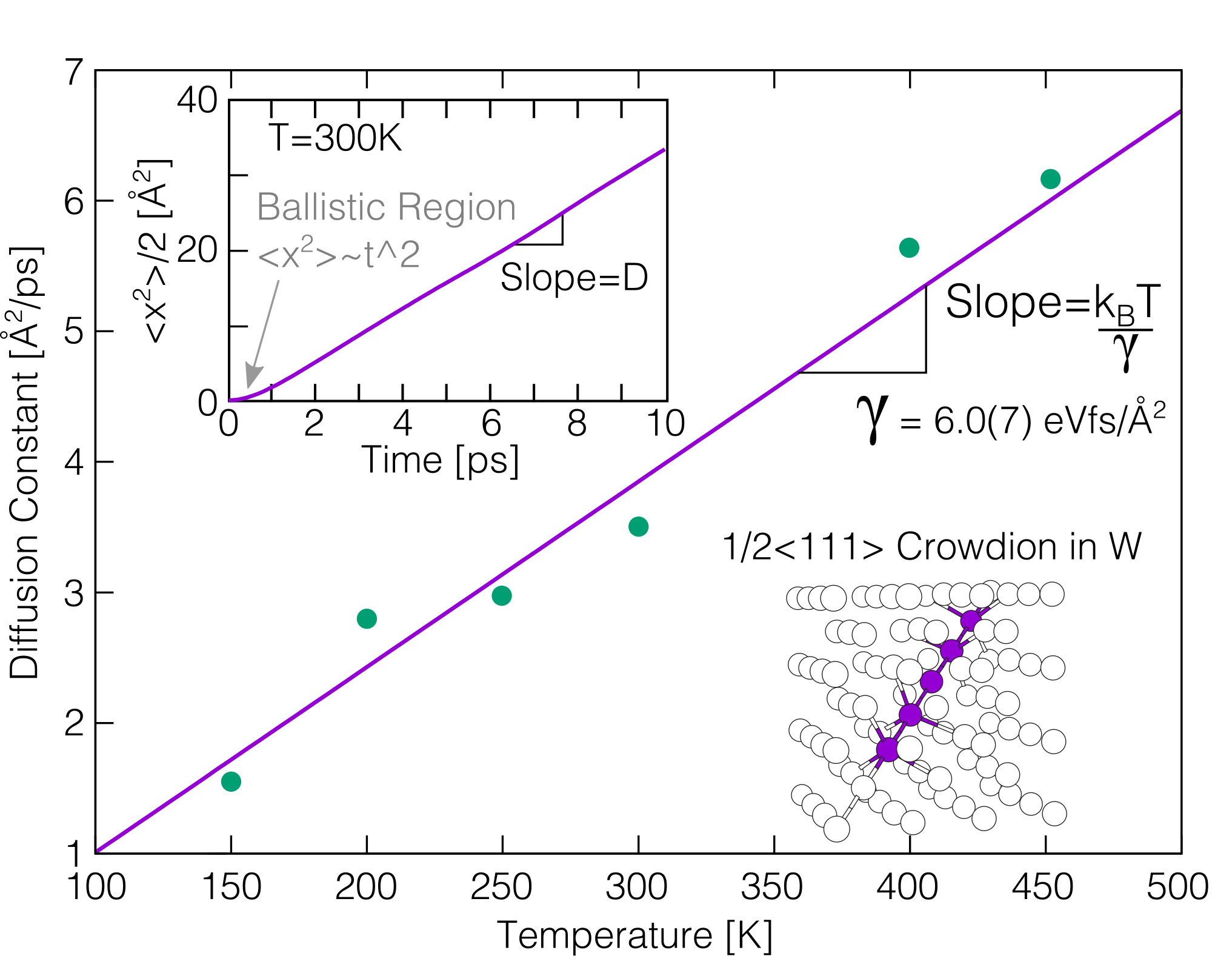}
\caption{The diffusivity of an interstitial crowdion defect in W at various temperatures. A linear fit gives $D={\rm k_BT}/\gamma_0$. Inset: typical data for the mean square displacement against time. We see ballistic behaviour at very short times, leading to diffusive behaviour after $\sim0.5$ ps.(Colour Online)\label{CR_D_PL}}
\end{figure}

The previous analytically tractable test case allowed a detailed analysis of our main result (\ref{full_gamma_0}), but the present work was motivated by the observation that the drag coefficient $\gamma\simeq\gamma_0$ is independent of temperature for many nanoscale defects in atomistic molecular dynamics (MD) simulations. For such nanoscale defects the drag coefficient is typically extracted under zero applied stress by calculating a diffusion coefficient\cite{zepeda2005,Dudarev2011,swinburne2013} from the defect trajectory. It is well known that for a particle of mass $\tilde{m}$ at temperatures above any migration barriers present we have the Ornstein - Uhlenbeck relation\cite{coffey2004}
\begin{equation}
\frac{\langle x^2(t)\rangle}{2}=\frac{\rm k_BT}{\gamma}t\left[1-\exp\left(-\frac{\gamma}{\tilde{m}}t\right)\right],
\end{equation}
that shows the transition from ballistic behaviour $\langle x^2(t)\rangle/t^2\sim\tilde{m}{\rm k_BT}$ at short times to diffusive behaviour at long times with the Einstein relation\cite{reif}
\begin{equation}
D=\lim_{t\to\infty}\frac{\langle x^2(t)\rangle}{2t}=\frac{\rm k_BT}{\gamma}.\label{D_PARAM}
\end{equation}
Equation (\ref{D_PARAM}) will be the benchmark value for $\gamma$ against which we test our results. In this paper, we have simulated a $1/2\langle111\rangle$ interstitial crowdion defect in Tungsten, which is known to posses an anomalous temperature independent drag coefficient $\gamma=\gamma_0$\cite{Dudarev2008b}. The simulations were performed with the LAMMPS package\cite{LAMMPS} using an EAM potential\cite{daw1984} for W by Marinica {\it et al.}\cite{marinica2013}. For the calculation of $D$ from (\ref{D_PARAM}) we used supercells of $\sim50,000$ atoms ($30\times 30 \times 30$ unit cells) which were relaxed, thermalized using a Langevin thermostat, and then run for 1-2ns over a range of temperatures to extract a diffusion constant $D$, as shown in figure \ref{CR_D_PL}. As $D$ was observed to rise linearly with temperature, we performed a least squares fit and used (\ref{D_PARAM}) to find a value of
\begin{equation}
\gamma=6.0\;(7)\;\; {\rm eV\cdot fs}/{\rm\AA}^2,
\end{equation}
where units are a reflection of the fact that $\gamma$ represents the impulse density of a heat bath. We also note that with an effective mass $\tilde{m}\sim m/7$, we can predict a `ballistic-diffusive transition' time-scale of
\begin{equation}
\frac{\tilde{m}}{\gamma}\simeq 0.43\;(5)\;\; {\rm ps},
\end{equation}
which is entirely consistent with the $\langle x^2(t)\rangle$ data in figure \ref{CR_D_PL}, showing that the overdamped Langevin equation (\ref{eom_od}) can adequately capture the stochastic dynamics of crystal defects on timescales larger than a picosecond.

\subsection{Evaluation of $C(s)$ in Molecular Dynamics}
We have developed a numerical scheme to determine $\partial_{\rm r}{\bf U}$ in dynamical MD simulation, allowing us to calculate the defect force
\begin{equation}
-\partial_{\rm r}V(t)=
-\partial_{\rm r}{\bf U}\cdot{\bm\nabla}V(t)
\end{equation}
which can then be used to calculate the force autocorrelation $C(s)$. From the analysis of section \ref{sec:ZW} we derived the result
\begin{equation}
\gamma=\int_0^\infty C(s){\rm d}s,
\end{equation}
thereby allowing us to calculate $\gamma$ from the trajectory of the defect force. The scheme is as follows. We first use the stable crystal defect structure at zero temperature to define the defect configuration ${\bf U}(\rm r)$, using the nudged elastic band (NEB) barrier climbing technique\cite{neb} to generate ${\bf U}(\rm r)$ for all values of the defect position within a lattice period parallel to the defect motion. All other lattice periodic images may then be generated simply by rigid translation. For the simple interstitial defect considered here we have found that to an extremely good approximation the set of NEB configurations may be generated from a single ground state configuration, which can offer computational efficiencies in the dynamical implementation of the method.\\

As detailed in appendix \ref{app:du}, as the deviation of ${\bf U}(\rm r)$ from the configuration of a perfect lattice varies slowly across atomic sites, we can equate $\partial_{\rm r}{\bf U}(\rm r)$ to the atomistic strain field along the direction of defect motion. We can then use linear interpolation to evaluate ${\bf U}(\rm r)$ and $\partial_{\rm r}{\bf U}(\rm r)$ for all values of the defect position $\rm r$ within a lattice period parallel to the defect motion. The application of this method to extended defects will be presented in detail elsewhere, though we have also found good results for certain glissile edge dislocation lines\cite{swinburne2014} and we expect this interpolation technique to be valid for edge-type defects with wide cores, as the atomic displacements are in the direction of defect motion and vary slowly with defect position. \\

Once ${\bf U}(\rm r)$ and $\partial_{\rm r}{\bf U}$ have been calculated, we run an ensemble of finite temperature, zero-stress MD simulations of the same system extracting the 3N-dimensional crystal configuration ${\bf X}(t)$, velocity $\dot{\bf X}$ and force $-{\bm\nabla}V({\bf X}(t))$ under micro-canonical conditions\cite{swinburne2013}. Whilst in principle we can extract this data every MD timestep (here 1fs), we have found that data can be taken every $5-10$ MD timesteps, still capturing the fastest vibrational modes in the force autocorrelation ($\sim50$fs) whilst gaining some efficiency from time coarse-graining the subsequent data analysis. For each extracted configuration we find, to within a small tolerance, the zero temperature configuration ${\bf U}(\rm r)$ that minimizes the quadratic weight $|({\bf X}-{\bf U}(\rm r))\cdot\partial_{\rm r}{\bf U}|^2$.\\

\begin{figure}
\centering
\includegraphics[width=0.48\textwidth]{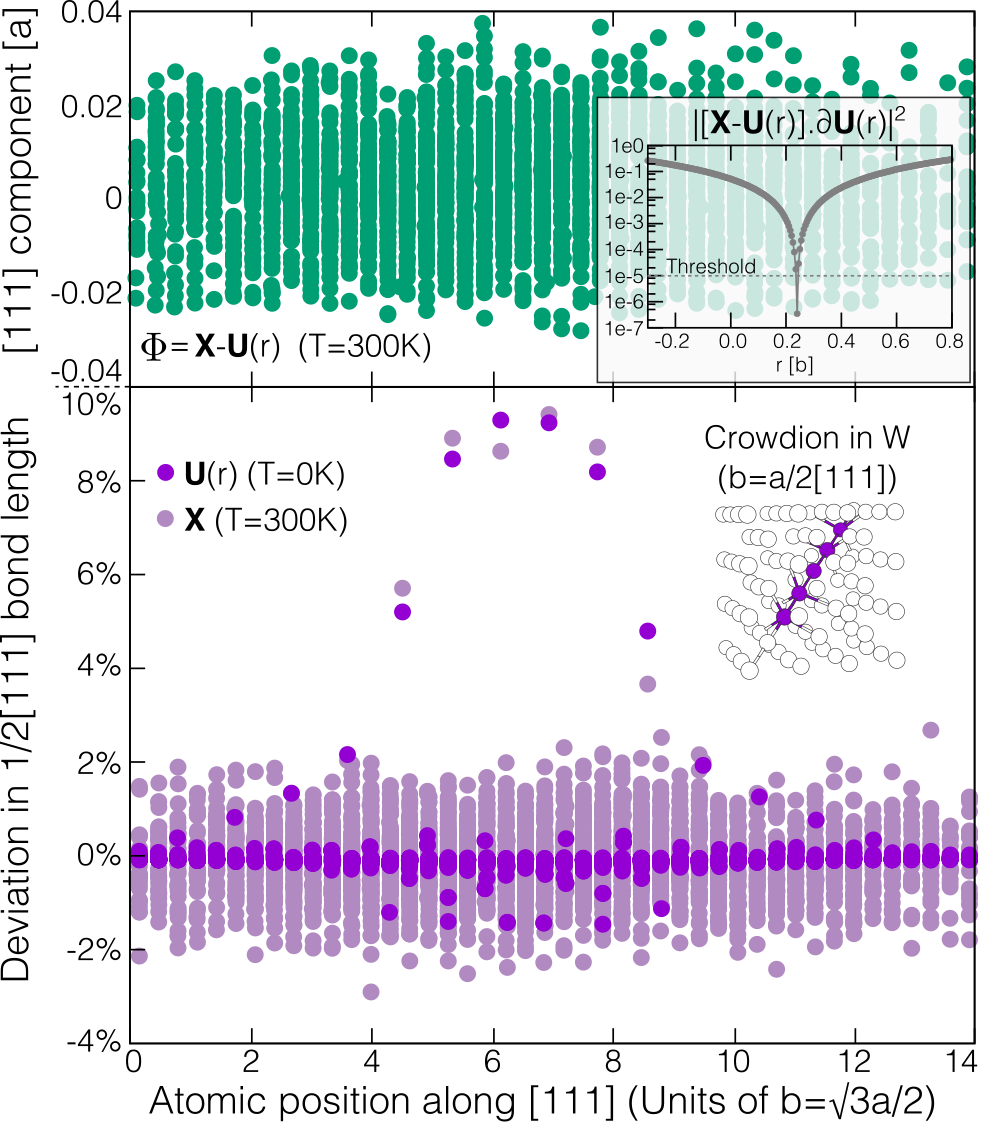}
\caption{Determination of ${\bf U}(\rm r)$ from MD simulation of a $1/2\langle111\rangle$ crowdion in W at T=300K. Below: The deviation in $1/2[111]$ bond length for ${\bf U}(\rm r)$ and ${\bf X}(t)$. Inset: illustration of a $1/2\langle111\rangle$ crowdion from\cite{Dudarev2008}. Above: The thermal vibration vector ${\bm\Phi}={\bf X}(t)-{\bf U}({\rm r})$, which fluctuates around zero with no peaks, as expected. Inset: Logarithmic plot of the quadratic weight $|({\bf X}-{\bf U}(\rm r))\cdot\partial_{\rm r}{\bf U}|^2$ for various values of $\rm r$. We see a quadratic minimum of $\sim10^{-7}$ which may be readily detected. (Colour Online)\label{TA_dU}}
\end{figure}

The choice of the quadratic weight $|({\bf X}-{\bf U}(\rm r))\cdot\partial_{\rm r}{\bf U}|^2$ is important, as its minimum should ideally vanish, as $|({\bf X}-{\bf U}(\rm r))\cdot\partial_{\rm r}{\bf U}|^2=|{\bm\Phi}\cdot\partial_{\rm r}{\bf U}|^2=0$, regardless of the defect structure or temperature. We find that this can be achieved to a very good approximation at finite temperature, as shown in the inset of figure \ref{TA_dU}; the minimum value of the quadratic weight $|[{\bf X}-\partial_{\rm r}{\bf U}]\cdot\partial_{\rm r}{\bf U}|^2$ is typically $\sim10^{-7}$, meaning a threshold value of $|({\bf X}-{\bf U}(\rm r))\cdot\partial_{\rm r}{\bf U}|^2\lesssim10^{-5}$ gives satisfactory accuracy in the defect position. Figure \ref{TA_dU} demonstrates that after minimising $|[{\bf X}-{\bf U}(\rm r)]\cdot\partial_{\rm r}{\bf U}|^2$ as described above, any crystal configuration ${\bf X}(t)$ can be split into a defect configuration ${\bf U}(\rm r)$ and a featureless, fluctuating field of thermal vibrations $\bm\Phi$. This is an important justification of the claim (\ref{first}) made at the beginning of section \ref{sec:PO} that such a decomposition was possible at finite temperatures.\\

\begin{figure}
\centering
\includegraphics[width=0.48\textwidth]{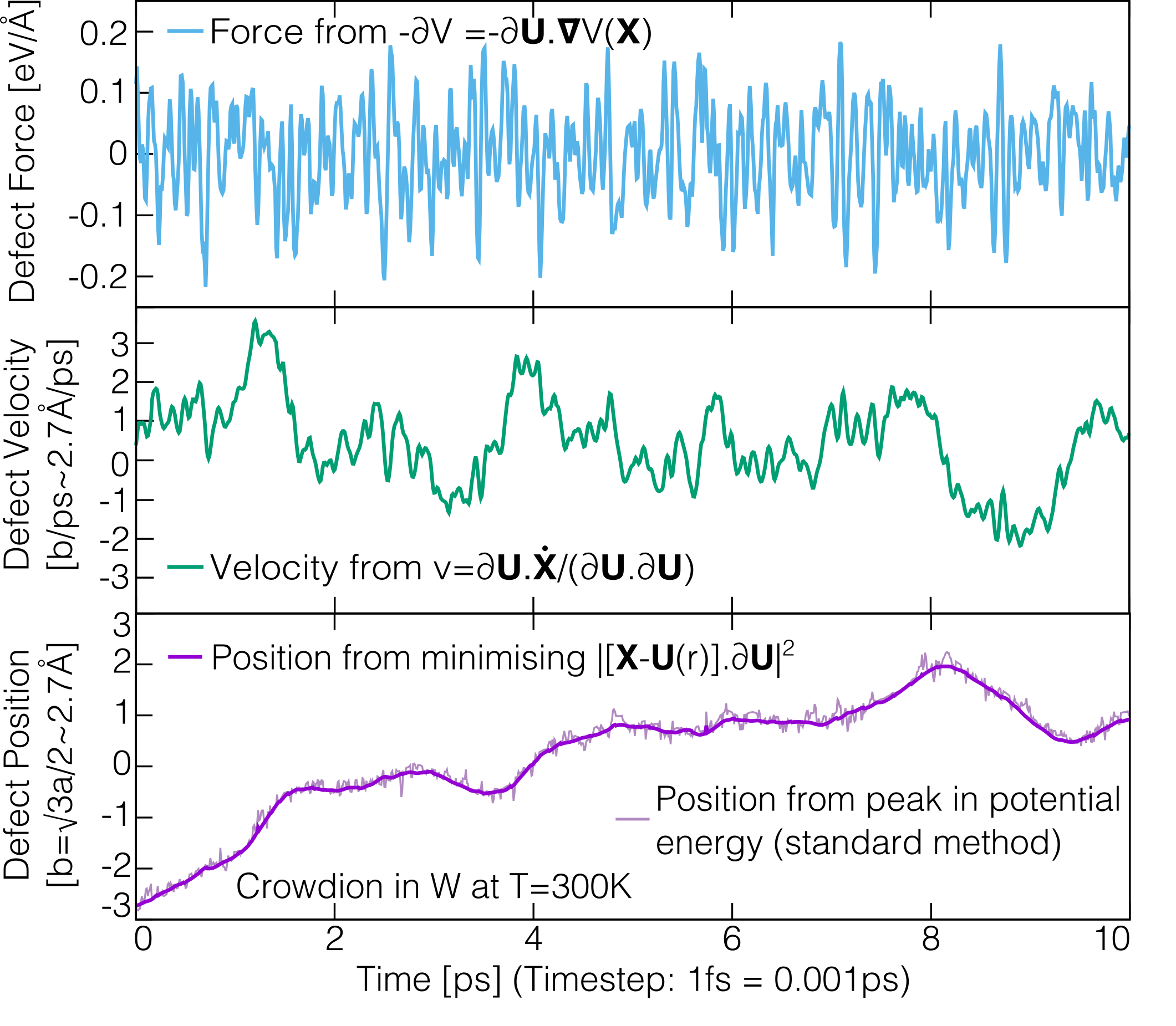}
\caption{The defect position, velocity and force acting on the defect, extracted for an interstitial crowdion in W at T=300K. To our knowledge this is the first time a defect velocity and force have been extracted directly from the velocity and force vectors of a simulated crystal. (Colour Online)\label{PVF}}
\end{figure}

We have found that the determination of ${\bf U}(\rm r)$ at each time step can be significantly accelerated by using the identity $m\partial_{\rm r}{\bf U}\cdot\dot{\bf X}(t)=\tilde{m}{\rm v}(t)$ (see figure \ref{PVF}) to approximate ${\rm r}(t+\delta t)$ by ${\rm r}(t)+\delta t{\rm v}(t)$. We have found this effective preconditioning of the minimisation condition means a satisfactory minimum may be found extremely quickly, with our method becoming comparable in speed to the typical method of extracting a defect position through analysis of the atomic disregistry. The efficacy of this method is a confirmation of our earlier assumption that the defect position typically varies very little over a thermal vibration time scale of $\sim0.1$ps $\sim100$ MD timesteps.\\

As shown in figure \ref{PVF}, the defect position trajectory obtained in this manner is typically much smoother than those found by analysing peaks in the potential energy or centrosymmetry parameter, as we fit a rigid defect profile to the whole configuration, removing fluctuations from the fitting procedure. However, the real benefit of our approach is that we may use $\partial_{\rm r}{\bf U}$ to project out the defect velocity ${\rm v}=\partial_{\rm r}{\bf U}\cdot\dot{\bf X}(t)/(\partial_{\rm r}{\bf U}\cdot\partial_{\rm r}{\bf U})$ and the force $-\partial_{\rm r}V=-\partial_{\rm r}{\bf U}\cdot{\bm\nabla}V({\bf X}(t))$ acting on a defect at each timestep. To the best of our knowledge this is the first time such quantities have been directly extracted from the atomic forces and velocities in an MD simulation.\\

The defect velocity is important as it can be used to aid the fitting procedure as described above, but it also gives important data to test the assumptions in our approach. We have calculated the trajectory average $\langle\rm v^2\rangle$ over a variety of temperatures and found that it gives the equipartition value $\tilde{m}{\rm k_BT}$ to within 10\%. Furthermore, the root mean velocity is typically 1 or 2 Burgers vectors b per ps, where $b=\sqrt{3}a/2$ is the translational period of the defect, meaning that over a vibrational period $\sim0.1$ ps the defect center moves at most 10\% of a migration period, supporting the timescale separation argument that we made at the start of section \ref{sec:ZW}. The defect force is, as expected, a rapidly fluctuating function of time with zero mean.

\begin{figure}
\includegraphics[width=0.48\textwidth]{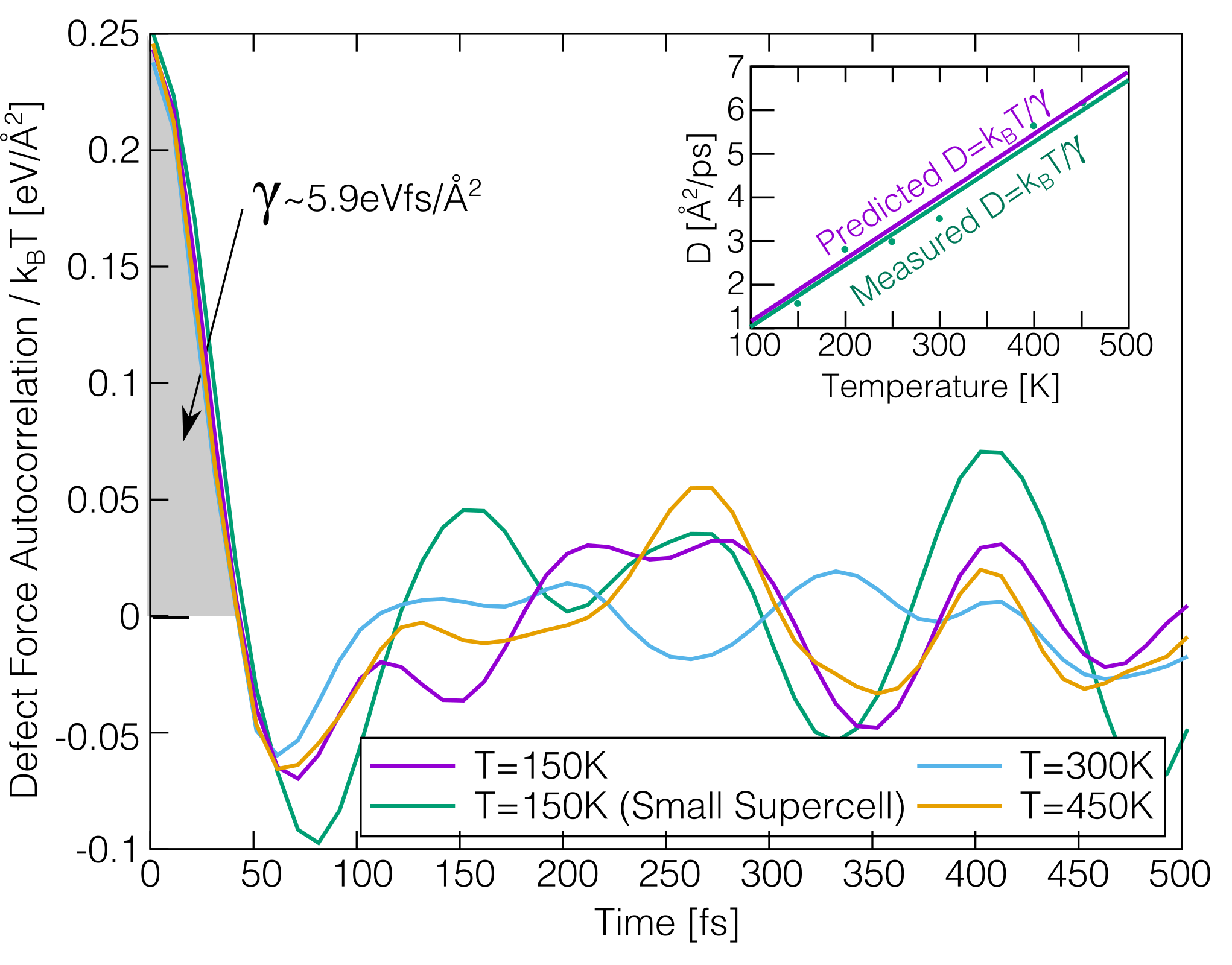}
\caption{The defect force autocorrelation calculated from molecular dynamics simulations at various temperatures. The small supercell used for one of the T=150K autocorrelations contained $\sim3,000$ atoms, whilst the other data was taken from supercells containing $\sim10,000$ atoms. We see the initial peak of the force autocorrelation divided by $\rm k_BT$ is essentially independent of temperature, giving an estimate for a temperature independent drag parameter $\gamma=\gamma_0\simeq5.9$eV$\cdot$ fs/${\rm\AA}^2$ that compares well with the value of $\gamma_0$=6.0(7) eV$\cdot$ fs/${\rm\AA}^2$ obtained from measurement of the diffusion constant $D={\rm k_BT}/\gamma_0$ (inset). (Colour Online)\label{FAC_CR}}
\end{figure}

With a defect force trajectory it is now a simple matter to calculate the defect force autocorrelation, which is then divided by $\rm k_BT$ to produce the memory kernel $C(s)$. From equation (\ref{GAMMA_MARK}), the time integral of $C(s)$ should be equal to the defect drag parameter $\gamma$. As shown in figure \ref{FAC_CR}, the initial peak of $C(s)$ is essentially identical across a wide range of temperatures, with the subsequent signal oscillating around zero. Importantly, we see that $C(s)$ decays to zero after $\sim0.05$ps$\sim\tau_{\rm D}/2$ over which time the defect is observed (figure \ref{PVF}) to be essentially stationary, again validating the assumption of timescale separation we made at the start of section \ref{sec:AG}. As the signal varies significantly and incoherently across different simulations, and typically flattens as we increase the simulation time and system size, we limit the time integration to the first zero in the memory kernel, finding a value for $\gamma=\gamma_0$ that is in good agreement with $\gamma$ extracted from analysis of $D={\rm k_BT}/\gamma$ from long diffusive trajectories generated using larger simulation supercells, typically containing $50,000$ atoms, as compared with $10,000$ atoms used for producing the force autocorrelation data. It is interesting to note that the peak of the force autocorrelation does not significantly vary across systems sizes and even for low temperatures (T=50K) where the very small lattice migration barrier $\sim0.01$eV $\sim100{k_B}$ causes the defect not to migrate only very rarely over a simulation time of a nanosecond. The precise nature of these finite size effects and their role in data analysis is a subject of a separate study.
\subsection{Evaluation of $C(s)$ in Molecular Statics}
In section \ref{sec:AG} we derived an analytical expression ({\ref{full_gamma_0}) for the temperature independent $\gamma_0$ in the form $\gamma_0=\int_0^\infty C_0(s){\rm d}s$, where the temperature independent memory kernel $C_0(s)$ read
\begin{equation}
C_0(s)=\sum_q\frac{\left(\partial_{\rm r}{\bf U}\cdot{\bm\nabla}^2V\cdot{\bf e}_q\right)}{m\omega^2_q}\cos(\omega_qs).\label{fac_0}
\end{equation}
The summation in (\ref{fac_0}) is over the set $\{{\bf e}_q,\omega_q\}_{q=1}^{3N-1}$ of vibrational eigenvectors and eigenfrequencies of the $\bm\Phi$- Hessian
\begin{equation}
{\bm\nabla}^2_{\bm\Phi}V=
\left(\mathbb{I}-\frac{\partial_{\rm r}{\bf U}\otimes\partial_{\rm r}{\bf U}}{
\partial_{\rm r}{\bf U}\cdot\partial_{\rm r}{\bf U}
}\right)
\cdot{\bm\nabla}^2V\cdot
\left(\mathbb{I}-\frac{\partial_{\rm r}{\bf U}\otimes\partial_{\rm r}{\bf U}}{\partial_{\rm r}{\bf U}\cdot\partial_{\rm r}{\bf U}}\right).
\end{equation}
As this expression only involves the calculation of second order derivatives it can be readily evaluated from a zero temperature configuration. After minimising the system to obtain ${\bf U}({\rm r})$, it is a simple matter to construct the matrix
\begin{equation}
\mathbb{I}-\frac{\partial_{\rm r}{\bf U}\otimes\partial_{\rm r}{\bf U}}{\partial_{\rm r}{\bf U}\cdot\partial_{\rm r}{\bf U}}
\end{equation}
which spans the $3N-1$ directions orthogonal to $\partial_{\rm r}{\bf U}({\rm r})$. We can then evaluate the Hessian matrix ${\bm\nabla}^2V$ of second derivatives (see appendix \ref{app:ten}) at ${\bf X}={\bf U}({\rm r})$ to construct the $\bm\Phi$-Hessian ${\bm\nabla}^2_{\bm\Phi}V$, defined in equation (\ref{VIB_HESS}). Using standard LAPACK routines, we can then obtain the $3N-1$ eigenvectors and eigenvalues $\{{\bf e}_q,m\omega^2_q\}_{k=1}^{3N-1}$, which may be used to evaluate (\ref{fac_0}). The result of these calculations are shown in Fig. \ref{FAC_TH_CR}. We see that the initial peak of the calculated force autocorrelation is in good agreement with the dynamical measurements and varies little when using a tiny system of 433 atoms ($6\times6\times6$ supercell) compared to a larger system of 3457 atoms ($12\times12\times12$ supercell). The slight additional amplitude of the force autocorrelation for the larger cell can be explained by analysing the coupling $(\partial_{\rm r}{\bf U}\cdot{\bm\nabla}^2V\cdot{\bf e}_q)/{m\omega^2_q}$ of the defect to each vibrational mode, shown in the inset of figure \ref{FAC_TH_CR}. We see a remarkably similar profile in the number of modes that couple strongly to the defect, but that a larger system clearly has a greater number of modes that couple only weakly, which will all contribute a small amount to the summation in (\ref{full_gamma_0}). The study of this vibrational coupling for a variety of defects will be the topic of a future work, but we finish this section by noting that our central result, the expression (\ref{fac_0}) for the temperature independent memory kernel $C_0(s)$, giving $\gamma_0=\int C_0(s){\rm d}s$, closely resembles the famous Kac-Zwanzig memory kernel $C_{\rm KZ}(s)$ for a particle connected to a bath of harmonic oscillators.  The Kac-Zwanzig memory kernel reads\cite{Zwanzig}
\begin{equation}
C_{\rm KZ}(s)=\sum_q\frac{\lambda}{m\omega^2_q}\cos(\omega_qs),\label{fac_zw}
\end{equation}
which is identical to $C_0(s)$ except that here the coupling constant $\lambda$ is a constant as opposed to $\left(\partial_{\rm r}{\bf U}\cdot{\bm\nabla}^2V\cdot{\bf e}_q\right)$, and the normal modes are typically assumed to be that of a continuous, homogeneous three-dimensional medium, i.e. $\omega=ck$, with $\sim\omega^2$ of oscillators of frequency $\omega$, meaning the memory kernel (\ref{fac_zw}) becomes precisely a delta function. It is interesting that crystal defects have, in realistic systems, a very heterogeneous coupling to their environment, but nevertheless the total coupling strength varies only a little as we consider larger systems.

\begin{figure}
\includegraphics[width=0.48\textwidth]{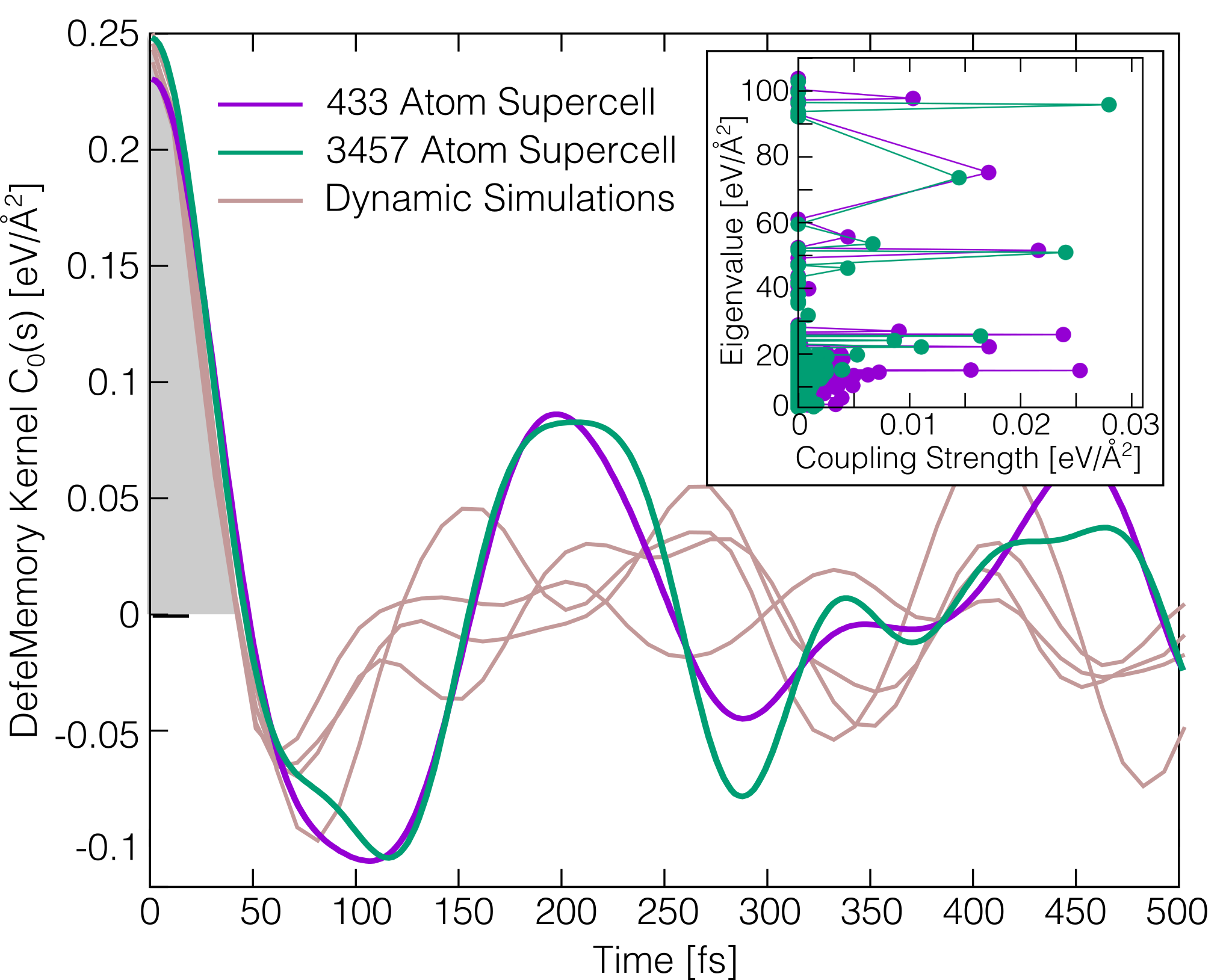}
\caption{The defect force autocorrelation calculated from molecular dynamics simulations at various temperatures. We see that the autocorrelation is independent of temperature, giving an estimate for a temperature independent drag parameter $\gamma=\gamma_0$ that can accurately predict the diffusion constant $D={\rm k_BT}/\gamma_0$ (inset). (Colour Online)\label{FAC_TH_CR}}
\end{figure}

\section{Conclusions}
In this paper we have developed a general method to define, construct and evaluate the stochastic equation of motion for a crystal defect. By defining a crystal defect only as a localised deformation of a non-linear, discrete crystal of $N$ atoms, we were able to identify a $3N$ dimensional vector $\partial _{\rm r}{\bf U}$ that gave the direction of defect motion in the configurational space of the crystal, i.e. the set of $N$ atomic displacements that effect defect migration from a given defect position. This vector is readily evaluated {\it in silico} and may be used to extract a defect position, velocity, and force directly from the $3N$ dimensional atomic positions, velocities and forces used in any MD simulation. This allows one to analyse much more than simply the defect trajectory with time and gives real fundamental insight into the defect dynamics.\\

The vector $\partial _{\rm r}{\bf U}$  allowed us to define the set of all $3N$-1 mutually orthogonal directions as a subspace of thermal vibrations consistent with a given defect position, with an equation of motion for all of these quantities emerging from the atomic equations of motion. Using the Mori-Zwanzig technique, we then derived, under well defined approximations, a Green-Kubo relation (\ref{GAMMA_MARK}) for the defect drag parameter $\gamma$ that equates $\gamma$ to the time integral of the defect force autocorrelation divided by temperature. A numerical implementation found that this method agreed well with traditional estimates of the diffusivity using the position trajectory, but that high quality data on the force autocorrelation could be obtained, due to the much smaller correlation time, from much smaller simulation supercells, offering an interesting method to extract long time behaviour from a relatively short dynamical simulation.

Under the assumption of a well defined system temperature (Gaussian Gibbs distribution) we also derived an analytical result (\ref{full_gamma}) for $\gamma$ in terms of the system potential energy. We found that in general $\gamma=\gamma_0+{\rm k_BT}\gamma_{\rm w}$, where $\gamma_0$ is a temperature independent term that is forbidden to exist in all previous theoretical treatments based on phonon scattering formalism, but has been widely observed in MD simulations. We found that $\gamma_0$ arises because the vibrations {\it orthogonal to defect motion} are in general not eigenmodes of the crystal. This was shown to violate a founding assumption of phonon scattering calculations, namely that thermal vibrations may be considered as the normal modes of a perfect crystal, which then interact with crystal defects through the introduction of {\it anharmonic} terms in the potential energy.

In a real crystal, the presence of a defect drastically changes the normal mode structure, and in general this new set of normal modes will not be orthogonal to the direction of defect motion, meaning that the vibrations ${\bf e}_q$ will couple to the defect even to quadratic order, i.e. through the quadratic coupling term ${\bf e}_q\cdot{\bm\nabla}^2V\cdot\partial_{\rm r}{\bf U}\neq0$.\\

To test this conclusion we investigated an extreme case, the integrable sine-Gordon system, where analytical results predict that in the presence of a soliton the normal modes {\it are} all orthogonal to the direction of defect motion, meaning in turn that for this integrable model $\gamma_0=0$. However, when simulating the sine-Gordon system, even with a very fine discretisation, we saw that whilst traditional {\it static} discreteness effects such as the Peierls barrier are effectively zero, $\gamma_0\neq0$ was still present, as fluctuations persist up to the smallest length scales in any discrete system, meaning they cannot be approximated away. As a result we expect the presence of $\gamma_0$ to be a general feature in the stochastic dynamics of any crystal defect.

Future work will concern detailing the application of this approach to extended defects such as dislocation lines and glissile stacking faults, and investigating how these techniques can be applied to develop a stochastic equation of motion for defects that posses a large migration barrier.

This work has been carried out within the framework of the EUROfusion Consortium and has received funding from the Euratom research and training programme 2014-2018 under grant agreement No 633053. The views and opinions expressed herein do not necessarily reflect those of the European Commission.
\appendix
\section{Calculation of $\partial_{\rm r}{\bf U}$ from a single atomic configuration}\label{app:du}
In the present work we focus on the case where defects posses very small migration barrier, which invariably means that the atomic displacements caused by the defect vary slowly along the direction of defect motion, as a small migration barrier implies no individual atom migrates a large distance under defect migration\cite{PeierlsOld,swinburne2013}. We now exploit this slow variation to show how one can calculate $\partial_{\rm r}{\bf U}({\rm r})$ from a single configuration ${\bf U}({\rm r})$. Let the defect move in a direction parallel to some lattice vector direction $\bf a$, typically the Burgers vector $\bf b$. Consider the positions of an individual atom ${\bf u}_i({\rm r})\in\mathbb{R}^3,i\in[1,N]$ and a neighbouring atom which we assign an index $n(i)\in[1,N]$ such that far from the defect ${\bf u}_{n(i)}({\rm r})-{\bf u}_{i}({\rm r})\to{\bf a}$. We recover the full 3N-dimensional configuration ${\bf U}({\rm r})$ through tensor addition, i.e. ${\bf U}({\rm r})={\oplus}_i{\bf u}_i({\rm r})$. By virtue of the translational symmetry of the crystal lattice we know that the displacement of atom $n(i)$ for a defect position ${\rm r}$ is equivalent to the displacement of atom $i$ for a defect position ${\rm r}-a$, namely ${\bf u}_{n(i)}({\rm r})={\bf u}_{i}({\rm r}-a)+{\bf a}$. For a slowly varying displacement field we may now expand ${\bf u}_{i}({\rm r}-a)$  to give ${\bf u}_{i}({\rm r}-a)\simeq{\bf u}_{i}({\rm r})-a\partial_{\rm r}{\bf u}_{i}({\rm r})$; combining these manipulations with the definition $\partial_{\rm r}{\bf U}({\rm r})=\oplus_i\partial_{\rm r}{\bf u}_i({\rm r})$ we finally obtain
\begin{equation}
\partial_{\rm r}{\bf U}({\rm r})\simeq\bigoplus_i\left({\bf u}_{n(i)}({\rm r})-{\bf u}_{i}({\rm r})-{\bf a}\right)/a.
\end{equation}
This result, valid when the atomic displacements vary slowly along the direction of migration, allows us to calculate $\partial_{\rm r}{\bf U}({\rm r})$ simply through calculating a finite difference derivative for each atom and its nearest neighbour along $\bf a$.\\

\section{Equivalence of $-\partial_{\rm r}V_0$ to Eshelby's configurational force}\label{app:esh}
As in the treatment of the main text, we consider a crystal at zero temperature containing a defect which may be described by a single position parameter ${\rm r}$, such that the atomic configuration ${\bf X}={\bf U}({\rm r})\in\mathbb{R}^{3\rm N}$. In the presence of weak external tractions $\bf T\in\mathbb{R}^{3\rm N}$ which are assumed to be non-zero only far from the defect core, one can show that the defect force is precisely the configurational force on a crystal defect first derived by Eshelby\cite{eshelby1951}. As the applied tractions are weak, the total potential energy may be written as
\begin{equation}
V_0({\rm r})+{\bf T}\cdot{\bf U}({\bf r})\label{tot_E_PK},
\end{equation}
where the subscript $V_0$ indicates that ${\bf X}={\bf U}({\rm r})$. We can then be vary (\ref{tot_E_PK}) with respect to ${\rm r}$ to obtain
\begin{equation}
-\partial_{\rm r}V_0={\bf T}\cdot\partial_{\rm r}{\bf U}({\rm r}).
\end{equation}
The requirement that $\bf T$ is weak and applied far from the defect core is to ensure the linearity in $\bf T$ of the total energy (\ref{tot_E_PK}); if the perturbation was non-linear the continuously parametrized set of minimum energy configurations ${\bf U}({\rm r})$ would not be a global minimum in the presence of an external traction. We note that equivalence between elasticity and a fully non-linear discrete treatment is only expected to apply in this regime. To explicitly apply a surface traction, let $\bf T$ represent a force of $\pm A\hat{\bf n}\cdot{\bm\sigma}\in\mathbb{R}^3$ per atom for two bounding planes $\Sigma_\pm$ far from the defect core, where $A$ is the area per atom. The defect force is now
\begin{equation}
-\partial_{\rm r}V_0=A\sum_{{\bf u}_i\in\Sigma_+,\Sigma_-}
\hat{\bf n}\cdot{\bm\sigma}\cdot\partial_{\rm r}{\bf u}_i({\rm r}),\label{disc_force}
\end{equation}
where $\partial_{\rm r}{\bf u}_i({\rm r})\in\mathbb{R}^3$, $i\in[1,N]$ are the individual components of $\partial_{\rm r}{\bf U}({\rm r})$ for each of the $N$ atoms in the system. To take a continuum limit we let the set of displacements ${\bf u}_i({\rm r})-{\bf x}^0_i$, where ${\bf x}^0_i$ is the nearest perfect lattice position, be interpolated by the continuum vector field of displacements ${\bf u}(x;{\rm r})$ to give
\begin{equation}
-\partial_{\rm r}V_0=\int_{x\in\Sigma_+,\Sigma_-}{\rm d}{S}\hat{\bf n}\cdot{\bm \sigma}\cdot\partial_{\rm r}{\bf u}(x),\label{eshel}
\end{equation}
in direct agreement with Eshelby's result. By virtue of global rotational symmetry we can always ensure ${\bf \sigma}$ is symmetric under permutation of indices, meaning we can always replace $\partial_{\rm r}{\bf u}(x)$ in (\ref{eshel}) by its symmetric component, the strain field ${\bm\epsilon}=\partial_{\rm r}{\bf u}(x)+(\partial_{\rm r}{\bf u}(x))^{\rm T}$. Under mild assumptions on the nature of the displacement field one may also continue Eshelby's derivation to obtain the Peach-Koehler dislocation force.
\begin{widetext}
\section{Tensorial derivatives of an embedded atom potential}\label{app:ten}
The potential energy for a set of atoms $\{{\bf x}^i\}$ interacting through an embedded atom potential are typically of the form
\begin{equation}
{U(\{{\bf x}^i\}) = U_1[\lambda]-U_{1/2}[\omega]\quad,\quad
U_C[\phi] = \sum_i \left(\sum_{j\neq i}\phi(r^{ij})\right)^C}\equiv\sum_i \Theta^C_i
,\label{potential}
\end{equation}
where $r^{ij}=|{\bf x}^i-{\bf x}^j|>0$ is the Euclidean distance between atoms $i$ and $j$, $\lambda$ is a pair potential term and $\omega$, the keystone of the embedded atom method, represents the electronic density. In practice these potential terms are neglected once $r^{ij}$ exceeds some cut-off radius $r_{\rm max}$.\\
\subsection{Derivatives of the pairing function $\phi$}
The embedded atom potential (\ref{potential}) is built from pair-potential functions $\phi(|{\bf x}_i-{\bf x}_j|)=\phi^{ij}$ between pairs $ij$. To simplify later notation, we now define the first and second derivatives, which by the translational invariance of the argument will have permutation symmetry in the cartesian directions $\alpha,\beta,\gamma,\epsilon \in (x,y,z)$-
\begin{align}
\chi^{ij}_\alpha &\equiv \frac{\partial\phi^{ij}}{\partial x^i_\alpha} = -\frac{\partial\phi^{ij}}{\partial x^j_\alpha}=\chi^{[ij]}_\alpha,
\quad
\Rightarrow
\frac{\partial\Theta^C_i}{\partial x^k_\alpha} = C\Theta^{C-1}_i\sum_m \chi^{[km]}_\alpha(\delta_{ik}+\delta_{im})\\
\Psi^{ij}_{\alpha\beta}
&\equiv
\frac{\partial\chi^{ij}_\alpha}{\partial x^i_\beta} = -\frac{\partial\chi^{ij}_\alpha}{\partial x^j_\beta} =   \frac{\partial\chi^{ji}_\alpha}{\partial x^j_\beta} = \Psi^{(ij)}_{(\alpha\beta)},
\quad
\Rightarrow \frac{\partial\chi^{[ij]}_\alpha}{\partial x^k_\beta} = (\delta_{ik}-\delta_{jk})\Psi^{(ij)}_{(\alpha\beta)}\\
\end{align}
where $[],()$ indicate the antisymmetric and symmetric permutation symmetry. Practically, the Cartesian derivatives $\chi,\Psi$ of $\phi(r)$ are evaluated in spherical polar coordinates though we omit these standard results as there is quite enough algebra already. To aid the following, we also define the `reduced' quantities
\begin{equation}
\bar\Theta^{C-n}_i \equiv \frac{C!}{(C-n)!}\Theta^{C-n}_i
\quad,\quad
\bar\chi^i_\alpha \equiv \sum_m\chi^{im}_\alpha
\quad,\quad
\bar\Psi^i_{\alpha\beta} \equiv \sum_m\Psi^{im}_{\alpha\beta}\quad\text{etc.}
\end{equation}
\[
\frac{\partial\bar\Psi^i_{\alpha\beta}}{\partial x^k_\gamma} = \delta_{ik}\bar\Upsilon^i_{\alpha\beta\gamma}-\Upsilon^{ik}_{\alpha\beta\gamma}\quad\text{etc.}
\]
\subsection{First and Second Derivatives}
With these definitions, we can immediately write
\begin{equation}
\frac{\partial U_C}{\partial x^i_\alpha} =
C\sum_m\Theta^{C-1}_m\frac{\partial\Theta_m}{\partial x^i_\alpha}
=
C\sum_m\Theta^{C-1}_m\sum_n \chi^{[in]}_\alpha(\delta_{mi}+\delta_{mn})
=
C\sum_{m}
\chi^{[im]}_\alpha
(
\Theta^{C-1}_i
+
\Theta^{C-1}_m
),
\end{equation}
\begin{align}
\frac{\partial^2 U_C}{\partial x^i_\alpha\partial x^j_\beta} &=
C\sum_{m}
\frac{\partial\chi^{[im]}_\alpha}{\partial x^j_\beta}
(
\Theta^{C-1}_i
+
\Theta^{C-1}_m
)
+
C(C-1)\sum_{m}
\chi^{[im]}_\alpha
(
\frac{\partial\Theta_i}{\partial x^j_\beta}
\Theta^{C-2}_i
+
\frac{\partial\Theta_m}{\partial x^j_\beta}
\Theta^{C-1}_m
)\nonumber\\
&=
C\sum_{m}
\Psi^{(im)}_{\alpha\beta}(\delta_{ij}-\delta_{mj})
(
\Theta^{C-1}_i
+
\Theta^{C-1}_m
)
\nonumber\\
&+
C(C-1)\sum_{mn}
\chi^{[im]}_\alpha\chi^{[jn]}_\beta(\delta_{ij}+\delta_{in})\Theta^{C-2}_i
+
\chi^{[im]}_\alpha\chi^{[jn]}_\beta(\delta_{mj}+\delta_{mn})\Theta^{C-2}_m\nonumber\\
\end{align}
Which in our reduced notation reads
\begin{align}
&\frac{\partial U_C}{\partial x^i_\alpha}
=
\bar\chi^i_\alpha\bar\Theta^{C-1}_+\sum_m\chi^{im}_\alpha\bar\Theta^{C-1}_i
,\nonumber\\
&\frac{\partial^2 U_C}{\partial x^i_\alpha\partial x^j_\beta}
=
-\Psi^{(ij)}_{\alpha\beta}(\bar\Theta^{C-1}_i+\bar\Theta^{C-1}_j)
+
\sum_m\chi^{im}_\alpha\chi^{jm}_\beta\bar\Theta^{C-2}_m
\nonumber\\
&+
\sum_{(p\bar\alpha,q\bar\beta)\in\mathbb{P}(i\alpha,j\beta)}\chi^{pq}_{\bar\alpha}\bar\chi^q_{\bar\beta}\bar\Theta^{C-2}_q
+
\delta_{ij}
\left(
\bar\Psi^i_{\alpha\beta}\bar\Theta^{C-1}_i
+
\sum_{m}\Psi^{im}_{\alpha\beta}\bar\Theta^{C-1}_m
+
\bar\chi^i_\alpha\bar\chi^i_\beta\Theta^{C-2}_i,
\right)
\end{align}
where the index-coordinate permutation sum is explicitly
\begin{equation}
(p\bar\alpha,q\bar\beta) = (i\alpha,j\beta) , (j\beta,i\alpha).
\end{equation}
As any analytic partial derivative must be invariant to the order of differentiation, it will be advantageous to group terms in higher order derivatives as sums over such index-coordinate permutations. The overbars on the Greek coordinate symbols are used to distinguish them from the coordinate symbols outside of such sums.
\end{widetext}
\end{document}